\documentclass[12pt]{article}
\usepackage{epsfig}
\usepackage{axodraw}

\begin{document}
\title{Form Factors and Semileptonic Decay
of $D_s^+\to \phi \bar{\ell}\nu$ From QCD Sum Rule}
\author{Dong-Sheng Du$^b$, Jing-Wu Li$^b$   and Mao-Zhi
Yang$^{a,b}$
 \thanks{Email address: duds@mail.ihep.ac.cn (D.S. Du), lijw@mail.ihep.ac.cn (J.W. Li),
  yangmz@mail.ihep.ac.cn (M.Z.Yang)}\\
{\small $a$ CCAST (World Laboratory), P.O. Box 8730, Beijing 100080, China;}\\
{\small $b$ Institute of High Energy Physics,
 P.O. Box 918(4), Beijing 100039, China}\footnote{Mailing address}
}
\date{\empty}
\maketitle

\begin{picture}(0,0)
       \put(340,230){BIHEP-TH-2003-30}
       \put(340,210){\bf }
\end{picture}

\begin{abstract}
We calculate $D_s^+\to \phi$ transition form factors $V$, $A_0$, $A_1$ and $A_2$,
and study semileptonic decay of  $D_s^+\to \phi \bar{\ell}\nu$
based on QCD sum rule method. We compare our results of the ratios
of $V(0)/A_1(0)$, $A_2(0)/A_1(0)$, $\Gamma_L/\Gamma_T$, and the
total decay branching ratio of  $D_s^+\to \phi \bar{\ell}\nu$ with
experimental data, and find that they are consistent.
\end{abstract}

\section*{1. Introduction}

Semileptonic decay of charm meson is important for studying strong
and weak interactions. It can be used to test techniques developed
for solving perturbative and nonperturbative problems in Quantum
Chromodynamics (QCD), and to extract elements of
Cabibbo-Kobayashi-Maskawa (CKM) matrix. Semileptonic decay is
simpler than hadronic decay of charm meson because leptons do not
involve strong interaction. The amplitude of semileptonic decay
can be decomposed into several transition form factors due to
Lorentz property of the hadronic matrix element. The form factors
include all the nonperturbative effects. Several methods can be
used to treat these problems, such as quark model, QCD sum rule
and Lattice QCD, among which, QCD sum rule and Lattice QCD are
based on the first principle of QCD.

The method of QCD sum rules \cite{SVZ} has been widely used in
hadronic physics since its establishment in the late 1970s. For
semileptonic decays of charm meson, $D^+\to \bar{K^0} e^+\nu_e$
was firstly studied in QCD sum rule method with three-point
correlation function \cite{AEK}. Several years later, QCD sum rule
method was extended to semileptonic decays of $B$ meson, $B\to
D(D^*)\ell \bar{\nu}$ \cite{BD} and $B\to \pi e\nu$ \cite{AAO}. In
these works, form factors $f_+(q^2)$ and $f_V(q^2)$ are calculated
at the point $q^2=0$, where $q^2$ is the momentum transfer
squared. For the whole physical region of $0\le q^2\le q^2_{max}$,
the form factors are either assumed to be pole dominance
$f(0)/(1-\frac{q^2}{m^2_{pol}})$, or a linear approximation was
used.  In Refs \cite{PBD,PB}, $D\to \bar{K^0} e^+\nu_e$,
$\bar{K^0}^* e^+\nu_e$ and $D\to \pi e\nu$, $\rho e\nu$ were
studied, where QCD sum rule method was extended to a very large
value of $q^2$ with a careful treatment of non-Landau-type
singularities. $D_s$ decays to $\eta$ and $\eta^\prime$ final
states were studied in \cite{pcde}.

In this work, we study $D_s^+\to \phi \bar{\ell}\nu$ in QCD sum
rule method. This decay mode has been measured in experiment long
time before \cite{exp1,exp2,exp3,exp4}. Now It is necessary to
analyze it theoretically. We calculate up to contributions of
operators of dimension 6 in the operator product expansion (OPE)
and keep the mass of s-quark. In our result, the large
contributions come from unit operator $I$ (result of perturbative
diagram ) and condensate of operators of dimension 3. Operator of
dimension 5 gives smaller contribution. The contributions of
operators of dimension 6 are negligible. When calculating
contribution of perturbative diagram and gluon condensate
(operator of dimension 4),  Cutcosky's rule has been used.
Therefore subtraction of continuum contribution is conveniently
performed not only for perturbative diagram but also for
contribution of gluon condensate. After some long steps of
calculation, we find the contributions of diagrams for the gluon
condensate cancel each other, so there is no gluon-condensate
contribution in $D_s^+\to \phi$ transition. This is our new
finding.

There are two independent Borel parameters $M_1^2$ and $M_2^2$ in
manipulating three-point correlation functions. In general, to
simplify the numerical analysis, a fixed ratio of $M_1^2/M_2^2$
was taken in recent references. In this work, we make the
numerical analysis in the whole region of independent $M_1^2$ and
$M_2^2$ to select the stability ``window", so it is different from
taking a fixed ratio of these two Borel parameters.

We calculate $D_s\to\phi$ transition form factors $V$, $A_0$,
$A_1$, $A_2$ and the branching ratio of $D_s^+\to \phi
\bar{\ell}\nu$. Our result of the ratios of $V(0)/A_1(0)$,
$A_2(0)/A_1(0)$, $\Gamma_L/\Gamma_T$ ($\Gamma_L$ and $\Gamma_T$
denote decay width of $D_s^+$ to $\phi$ meson in longitudinal and
transverse polarization, respectively), and the total branching
fraction are in agreement with experimental data.

Recently, just before this work is finished, we find that
$D_s^+\to \phi\bar{\ell}\nu$ was also calculated in
Ref.\cite{IBMN}. However, their analysis is very different from
ours. First, by carefully choosing the requirement that the double
Borel parameters $M_1^2$ and $M_2^2$ should not be too large for
keeping the continuum contribution small, and at the same time,
$M_1^2$ and $M_2^2$ should not be too small for keeping the
truncated OPE series effective, i.e., keeping the contributions of
higher dimension operators small, we get very different stability
``window" for the Borel parameters. Second, our results of the
transition form factors are different from theirs. Especially for
$A_2$, they got negative value, however, we get positive. Using
their values of the form factors, although one can get the total
branching ratio of $D_s^+\to \phi \bar{\ell}\nu$ to be compatible
with experimental result, the ratio of $\Gamma_L/\Gamma_T$ will be
too large. But in our case, $\Gamma_L/\Gamma_T=0.99\pm 0.43$ which
is consistent with world average $0.72\pm 0.18$.

The paper is organized as follows. In section 2, we briefly
introduce the QCD sum rule method used in this work. Section
3 is the calculation. Section 4 is the numerical analysis and
discussion. Section 5 is devoted to the summary.

\section*{2. The method}

To calculate the transition form factors of semileptonic $D_s$ meson decays,
the standard procedure in QCD sum rule method is to consider the
three-point correlation function defined as
\begin{equation}
\Pi_{\mu\nu}=i^2\int d^4 x d^4 y e^{ip_2\cdot x-ip_1\cdot y}
 \langle 0|T\{j^\phi_\nu(x) j_\mu (0)j_5^D(y)\} |0\rangle ,
\label{correlator}
\end{equation}
with the currents having the same quantum numbers as the relevant
mesonic states under consideration, which are defined by: 1) the
current of $D_s$ channel, $j_5^D(y)=\bar{c}(y)i\gamma_5 s(y)$; 2)
the current of weak transition: $j_\mu (0)=\bar{s}\gamma_\mu
(1-\gamma_5)c$; 3) the current of $\phi$ channel:
$j_\nu^\phi(x)=\bar{s}(x)\gamma_\nu s(x)$.  On one hand, inserting
a complete set of intermediate hadronic states into the
correlation function, and using the double dispersion relation,
one can express the correlation function in terms of a set of
hadronic states,
\begin{equation}
\Pi_{\mu\nu}=\int ds_1 ds_2\frac{\rho (s_1,s_2,q^2)}{(s_1-p_1^2)
 (s_2-p_2^2)},
 \label{c2}
\end{equation}
with
$$
\rho (s_1,s_2,q^2)=\sum_{XY}\langle 0|j_\nu^\phi |X\rangle \langle
   X|j_\mu |Y\rangle \langle Y|j_5^D |0 \rangle \delta (s_1-m_Y^2)
    \delta (s_2-m_X^2)\theta(p_X^0)\theta(p_Y^0),
$$
where $X$ and $Y$ denote the complete set of hadronic states of
$\phi$ and $D_s$ channels, respectively. $p_X$ and $p_Y$ are the
four-momentum of $X$ and $Y$ states, $s_1=p_Y^2$, $s_2=p_X^2$, and
$q=p_1-p_2$. Integrate over $s_1$ and $s_2$ in Eq.~(\ref{c2}), we
can obtain
\begin{equation}
\Pi_{\mu\nu}=\sum_{XY}\frac{\langle 0|j_\nu^\phi |X\rangle \langle
   X|j_\mu |Y\rangle \langle Y|j_5^D |0
   \rangle}{(m_Y^2-p_1^2)(m_X^2-p_2^2)} +\mbox{\small{continuum states}}.
\end{equation}
Separate the ground states of $D_s$ and $\phi$ channels
apparently, the above equation becomes
\begin{equation}
\Pi_{\mu\nu}=\frac{\langle 0|j_\nu^\phi |\phi\rangle \langle
   \phi|j_\mu |D_s\rangle \langle D_s|j_5^D |0
   \rangle}{(m_{D_s}^2-p_1^2)(m_{\phi}^2-p_2^2)} +
   \mbox{\small{higher resonances and continuum states}}.
\end{equation}

The weak transition matrix element $D_s\to\phi$ can be decomposed
as
\begin{eqnarray}
\langle\phi(\varepsilon,p_2)|j_\mu | D_s(p_1)\rangle
&=&\varepsilon_{\mu\nu\alpha\beta}\varepsilon^{*\nu}p_1^\alpha
p_2^\beta
\frac{2V(q^2)}{m_{D_s}+m_{\phi}} \nonumber \\
&&-i(\varepsilon^*_\mu-\frac{\varepsilon^*\cdot q}{q^2}
q_\mu)(m_{D_s}+m_{\phi})A_1(q^2) \\
&&+i[(p_1+p_2)_\mu -\frac{m_{D_s}^2-m_{\phi}^2}{q^2} q_\mu ]
\varepsilon^*\cdot q \frac{A_2(q^2)}{m_{D_s}+m_{\phi}} \nonumber\\
&&-i\frac{2m_\phi \varepsilon^*\cdot q}{q^2}q_\mu
A_0(q^2),\nonumber
\end{eqnarray}
where $q=p_1-p_2$. The vacuum-to-meson transition amplitudes can
be parameterized through defining the corresponding decay
constants,
\begin{eqnarray}
&&\langle 0|\bar{s}\gamma_\nu s|\phi\rangle =m_\phi f_\phi
  \varepsilon_\nu^{(\lambda)}, \nonumber \\
&&\langle 0|\bar{s}i\gamma_5 c|D_s\rangle =
\frac{f_{D_s}m_{D_s}^2}{m_c+m_s}.
\end{eqnarray}
Finally the correlation function can be expressed in terms of
meson decay constants and $D_s\to \phi$ transition matrix element,
\begin{eqnarray}
\Pi_{\mu\nu}=\frac{ m_\phi f_\phi
  \varepsilon_\nu^{(\lambda)}\langle
   \phi(\varepsilon_\nu^{(\lambda)},p_2)|j_\mu |D_s(p_1)\rangle
   f_D m_{D_s}^2}{(m_{D_s}^2-p_1^2)(m_{\phi}^2-p_2^2)(m_c+m_s)}\nonumber\\
    +   \mbox{\small{higher resonances and continuum states}}.
    \label{c3}
\end{eqnarray}

On the other hand, the correlation function of
Eq.~(\ref{correlator}) can be evaluated at negative values of
$p_1^2$ and $p_2^2$ by the operator-product expansion in QCD, in
which the time-ordered current operators in Eq.~(\ref{correlator})
is expanded in terms of a series of local operators with
increasing dimensions,
\begin{eqnarray}
&&i^2\int d^4x d^4 y e^{ip_2\cdot x-ip_1\cdot y}
 T\{j^\phi_\nu(x) j_\mu (0)j_5^D(y)\} \nonumber\\
 &=&C_{0\mu\nu} I +C_{3\mu\nu} \bar{\Psi}\Psi
    +C_{4\mu\nu} G^a_{\alpha\beta}G^{a\alpha\beta}
    +C_{5\mu\nu} \bar{\Psi}\sigma_{\alpha\beta}T^a G^{a\alpha\beta}\Psi
    \nonumber\\
  &~+&C_{6\mu\nu}
 \bar{\Psi}\Gamma \Psi \bar{\Psi}\Gamma^{\prime}\Psi+\cdots,
 \label{opef}
 \end{eqnarray}
where $C_{i\mu\nu}$'s are Wilson coefficients, $I$ is the unit
operator, $\bar{\Psi}\Psi$ is the local Fermion field operator of
light quarks, $G^a_{\alpha\beta}$ is gluon strength tensor,
$\Gamma$ and $\Gamma^{\prime}$ are the matrices appearing in the
procedure of calculating the Wilson coefficients. Sandwich the
left and right hand sides of Eq.~(\ref{opef}) between two vacuum
states, we get the correlation function in terms of Wilson
coefficients and condensates of local operators,
\begin{eqnarray}
\Pi_{\mu\nu}&=&i^2\int d^4x d^4 y e^{ip_2\cdot x-ip_1\cdot y}
 \langle 0|T\{j^\phi_\nu(x) j_\mu (0)j_5^D(y)\}|0\rangle \nonumber\\
 &=&C_{0\mu\nu} I +C_{3\mu\nu} \langle 0|\bar{\Psi}\Psi|0\rangle
    +C_{4\mu\nu} \langle 0|G^a_{\alpha\beta}G^{a\alpha\beta}|0\rangle
    +C_{5\mu\nu} \langle 0|\bar{\Psi}\sigma_{\alpha\beta}T^a G^{a\alpha\beta}\Psi|0\rangle
    \nonumber\\
  &~+&C_{6\mu\nu}\langle 0|
 \bar{\Psi}\Gamma \Psi \bar{\Psi}\Gamma^{\prime}\Psi|0\rangle +\cdots,
 \label{conden}
\end{eqnarray}
For later convenience, we shall reexpress the above equation. In
general, it can be expressed in terms of six independent Lorentz
structures
\begin{equation}
\Pi_{\mu\nu}=-f_0\varepsilon_{\mu\nu\alpha\beta}p_1^\alpha
p_2^\beta-i(f_1 p_{1\mu}p_{1\nu}+f_2 p_{2\mu}p_{2\nu}+f_3
p_{1\nu}p_{2\mu}+f_4 p_{1\mu}p_{2\nu}+f_5g_{\mu\nu}). \label{c4}
\end{equation}
Each $f_i$ includes perturbative and condensate contributions
\begin{equation}
f_i=f_i^{pert}+f_i^{(3)}+f_i^{(4)}+f_i^{(5)}+f_i^{(6)}+\cdots,
\end{equation}
where $f_i^{(3)}$, $\cdots$, $f_i^{(6)}$ are contributions of
condensates of dimension 3, 4, 5, 6, $\cdots$ in
Eq.~(\ref{conden}). In next section we can see that perturbative
contribution and gluon condensate contribution can be finally
written in the from of dispersion integration,
\begin{eqnarray}
 f_i^{pert}&=&\int d s_1
 d s_2\frac{\rho^{pert}_i(s_1,s_2,q^2)}{(s_1-p_1^2)(s_2-p_2^2)},
 \nonumber  \\
f_i^{(4)}&=&\int d s_1
 d s_2\frac{\rho^{(4)}_i(s_1,s_2,q^2)}{(s_1-p_1^2)(s_2-p_2^2)}.
 \nonumber  \nonumber
 \end{eqnarray}
We approximate the contribution of higher resonances and continuum
states as integrations over some thresholds $s_1^0$ and $s_2^0$ in
the above equations. Then equate the two representations of the
correlation function in Eq.~(\ref{c3}) and (\ref{c4}), we can get
an equation for the form factors. To improve such equation, we
make Borel transformation over $p_1^2$ and $p_2^2$ in both sides,
which can further suppress higher resonance contribution. The
definition of Borel transformation to any function $f(p^2)$ is
$$\hat{B}_{\left|\frac{}{}\right.p^2,M^2}f(p^2)=\lim_{\small\begin{array}{ll}& n\to\infty \\
    & p^2\to -\infty  \\&-p^2/n=
    M^2  \end{array} } \frac{(-p^2)^n}{(n-1)!}\frac{\partial ^n}{\partial (p^2)^n}
    f(p^2).$$
Some examples of Borel transformation is given in the following,
\begin{eqnarray*}
&&\hat{B}_{\left|\frac{}{}\right.p^2,M^2}\frac{1}{(s-p^2)^k}=
\frac{1}{(k-1)!}\frac{1}{(M^2)^k}e^{-s/M^2},\\
&&\hat{B}_{\left|\frac{}{}\right.p^2,M^2}(p^2)^k=0,~~~ \mbox{for
any}~~ k\ge 0.
\end{eqnarray*}

Equating the two representations of the correlation function,
subtracting the higher resonances and continuum contribution, and
performing Borel transformation in both variables $p_1^2$ and
$p_2^2$, we finally obtain the sum rules for the form factors,
\begin{eqnarray}
V(q^2)&=& \frac{(m_c+m_s)(m_{D_s}+m_\phi )}{2m_\phi f_\phi
 f_{D_s}m_{D_s}^2}e^{m_{D_s}^2/M_1^2}e^{m_{\phi}^2/M_2^2}M_1^2M_2^2
 \hat{B}f_0, \nonumber \\
A_1(q^2)&=& -\frac{(m_c+m_s)}{m_\phi f_\phi
 f_{D_s}m_{D_s}^2(m_{D_s}+m_\phi )}e^{m_{D_s}^2/M_1^2}e^{m_{\phi}^2/M_2^2}M_1^2M_2^2
 \hat{B}f_5, \nonumber \\
A_2(q^2)&=& \frac{(m_c+m_s)(m_{D_s}+m_\phi )}{m_\phi f_\phi
 f_{D_s}m_{D_s}^2}e^{m_{D_s}^2/M_1^2}e^{m_{\phi}^2/M_2^2}M_1^2M_2^2
 \frac{1}{2}\hat{B}(f_1+f_3), \label{formsr} \\
A_0(q^2)&=& -\frac{(m_c+m_s)}{2m_\phi^2 f_\phi
 f_{D_s}m_{D_s}^2}e^{m_{D_s}^2/M_1^2}e^{m_{\phi}^2/M_2^2}M_1^2M_2^2
 [\hat{B}(f_1+f_3)\frac{m_{D_s}^2-m_\phi^2}{2}\nonumber \\&~&+
 \hat{B}(f_1-f_3)\frac{q^2}{2}+f_5], \nonumber
 \end{eqnarray}
where $\hat{B} f_i$ denotes Borel transforming $f_i$ in both
variables $p_1^2$ and $p_2^2$, $M_1$ and $M_2$ are Borel
parameters. Because we have subtracted the higher resonance and
continuum contribution, now the dispersion integration for
perturbative and gluon condensate contribution should be performed
under the threshold,
\begin{eqnarray}
 f_i^{pert}&=&\int^{s_1^2} d s_1
 \int^{s_2^2} d s_2\frac{\rho^{pert}_i(s_1,s_2,q^2)}{(s_1-p_1^2)(s_2-p_2^2)},
 \nonumber  \\
f_i^{(4)}&=&\int^{s_1^2} d s_1
 \int^{s_2^2}d s_2\frac{\rho^{(4)}_i(s_1,s_2,q^2)}{(s_1-p_1^2)(s_2-p_2^2)}.
 \nonumber  \nonumber
 \end{eqnarray}

In the next section, we will explain the technique of calculating
the Wilson coefficients and given the resulted form of the sum
rules for the form factors.

\section*{3. The Calculation of the Wilson Coefficients}

In this work, we first calculate the Wilson coefficients in the
operator-product expansion \cite{reind}, then extract the relevant
terms $f_i$'s for the sum rules of the form factors in
Eq.~(\ref{formsr}). We will not present the result of each Wilson
coefficient here because their forms are very tedious. We only
give the results of the form factors according to the contribution
of each condensate.

\subsection*{3.1 The Calculation of the Perturbative part}

The diagram for the perturbative contribution is depicted in
Fig.1.  The leading order in $\alpha_s$ expansion is considered
here. This contribution amounts to Wilson coefficient $C_0$ in OPE
representation of the correlation function in Eq.~(\ref{conden}).
We can write down this amplitude (see Fig.1),
\begin{figure}[h]
\begin{center}
\begin{picture}(150,150)(-15,-25.98)
\DashArrowLine(-15,-25.98)(0,0){3} \ArrowLine(0,0)(40,69.28)
\ArrowLine(0,0)(40,69.28)\ArrowLine(40,69.28)(80,0)\ArrowLine(80,0)(0,0)
\DashArrowLine(80,0)(95,-25.98){3}\Photon(40,69.28)(40,94.28){1}{4}
\put(41,70){$o$}
\put(80.5,0.5){$x$}
\put(-5.5,0.5){$y$}
\put(-24,-16){$p_1$}
\put(95,-16){$p_2$}
\put(10,34){$c$}
\put(38,-10){$s$}
\put(64,34){$s$}
\end{picture}
\caption{\small Diagram for perturbative contribution.}
\end{center}
\end{figure}
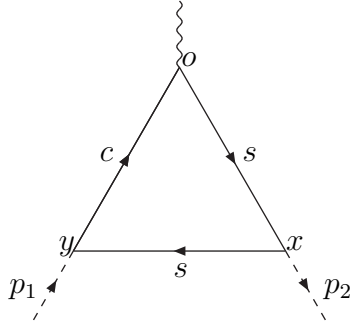

\begin{eqnarray}
C_0&=&i^2\int
\frac{d^4k}{(2\pi)^4}(-1)Tr\left[i\gamma_5\frac{i(\not{k}+m_s)}{k^2-m_s^2+i\varepsilon}
\gamma_\nu
\frac{i(\not{k}+\not{p}_2+m_s)}{(k+p_2)^2-m_s^2+i\varepsilon}\gamma_\mu
(1-\gamma_5)\right.\nonumber \\
&&\left.\frac{i(\not{k}+\not{p}_1+m_s)}{(k+p_1)^2-m_c^2+i\varepsilon}\right].
\label{pert1}
\end{eqnarray}
The above integration can be performed according to Cutkosky's
rule \cite{cutkosky}. That is, to write the integration of
Eq.~(\ref{pert1}) in the form of dispersion integration,
\begin{equation}
C_0=\int d s_1 d s_2
\frac{\rho(s_1^2,s_2^2,q^2)}{(s_1-p_1^2)(s_2-p_2^2)}.
\end{equation}
The spectral density $\rho(s_1,s_2,q^2)$ can be directly
calculated by substituting the denominators of the quark
propagators for $\delta$ functions, i.e., putting all the quark
lines on-mass-shell,
\begin{equation}
\frac{1}{k^2-m_s^2+i\varepsilon}\to -2\pi i \delta(k^2-m_s^2),
~\mbox{etc.},
\end{equation}
then the spectral density can be calculated from,
\begin{eqnarray}
\rho(s_1,s_2,q^2)&=&\frac{1}{(2\pi i)^2} (-2\pi i)^3
\int\frac{d^4k}{(2\pi)^4} Tr[\gamma_5(\not{k}+m_s)\gamma_\nu
(\not{k}+\not{p}_2+m_s)\gamma_\mu (1-\gamma_5)\nonumber \\
&&\times
(\not{k}+\not{p}_1+m_c)]\delta (k^2-m^2)\delta[(k+p_1)^2-m_1^2]\nonumber \\
&& \times\delta [(k+p_2)^2-m_2^2]_{\left|\small{\begin{array}{ll}
                                p_1^2\to s_1&\\
                                p_2^2\to s_2&
                                \end{array}}\right.} .
\end{eqnarray}
To perform the above integration, some basic formulas are needed.
Part of them have been given in Ref.\cite{ioffe} without the quark
mass, here we calculate them with the quark masses included,
\begin{eqnarray}
&&I=\int d^4k \delta (k^2-m^2)\delta[(k+p_1)^2-m_1^2]\delta
[(k+p_2)^2-m_2^2]=\frac{\pi}{2\sqrt{\lambda}},\label{i0}\\
&&I_\mu =\int d^4k k_\mu \delta
(k^2-m^2)\delta[(k+p_1)^2-m_1^2]\delta \
[(k+p_2)^2-m_2^2]\nonumber \\
&~~~&\equiv a_1 p_{1\mu}+b_1 p_{2\mu},\label{i1}\\[4mm]
&&\left\{ \begin{array}{ll}
          a_1~=&-\frac{\pi}{2\lambda ^{3/2}}[s_2(-s_1+s_2-q^2)+
            (s_1+s_2-q^2)(m^2-m_2^2)\\[3mm]
             &~-2s_2(m^2-m_1^2)],\\[3mm]
          b_1~=&-\frac{\pi}{2\lambda ^{3/2}}[s_1(-s_2+s_1-q^2)+
            (s_1+s_2-q^2)(m^2-m_1^2)\\[3mm]
              &~-2s_1(m^2-m_2^2)],
          \end{array}\right. \nonumber
\end{eqnarray}
\begin{eqnarray}
&&I_{\mu\nu}=\int d^4k k_\mu k_\nu\delta
(k^2-m^2)\delta[(k+p_1)^2-m_1^2]\delta \
[(k+p_2)^2-m_2^2]\nonumber
\\ &&\equiv a_2p_{1\mu}p_{1\nu}+b_2p_{2\mu\nu}+c_2(p_{1\mu}p_{2\nu}+
          p_{1\nu}p_{2\mu})+d_2g_{\mu\nu},\label{i2}\\[4mm]
&&\left\{ \begin{array}{ll}
      D_1~\equiv & s_1-m_1^2+m^2,~~~~~D_2\equiv s_2-m_2^2+m^2,\\[3mm]
      a_2~=& \frac{\pi}{\lambda^{3/2}}m^2s_2
        +\frac{1}{\lambda}[3s_2D_1a_1-(s_1+s_2-q^2)D_2b_1+s_2D_2b_1]\\[3mm]
      b_2~=& \frac{\pi}{\lambda^{3/2}}m^2s_1
        +\frac{1}{\lambda}[s_1D_1a_1-(s_1+s_2-q^2)D_1b_1+3s_1D_2b_1]\\[3mm]
      c_2~=&-\frac{\pi}{\lambda^{3/2}}m^2\frac{1}{2}(s_1+s_2-q^2)\\[3mm]
         &-\frac{1}{\lambda}[\frac{1}{2}(s_1+s_2-q^2)D_1a_1-2s_2D_1b_1
         +\frac{3}{2}(s_1+s_2-q^2)D_2b_1]\\[3mm]
      d_2~=&\frac{\pi}{4\sqrt{\lambda}}+\frac{1}{4}[D_1a_1+D_2b_1]
          \end{array} \right. ,\nonumber
\end{eqnarray}
where $\lambda (s_1,s_2,q^2)=(s_1+s_2-q^2)^2-4s_1s_2$, and in
Eqs.~(\ref{i0}) $\sim$ (\ref{i2}), substitutions $p_1^2\to s_1$
and $p_2^2\to s_2$ have been indicated.

\subsection*{3.2 The Contribution of bi-quark Operators $\bar{\Psi}(x)\Psi(y)$,
 $\bar{\Psi}(0)\Psi(x)$}

The diagrams for the contributions of $\bar{\Psi}(x)\Psi(y)$ and
$\bar{\Psi}(0)\Psi(x)$ are shown in Fig. 2. The contribution of
Fig. 2(b) is zero after double Borel transformation in both
variables $p_1^2$ and $p_2^2$ because only one variable appears in
the denominator $1/(p_2^2-m_s^2)$. So we will not consider Fig.
2(b) in the following. The contribution of Fig. 2(a) to the
correlation function is
\begin{figure}[h]
\begin{center}
\begin{picture}(300,150)(-15,-25.98)
\DashArrowLine(-15,-25.98)(0,0){3} \ArrowLine(0,0)(40,69.28)
\ArrowLine(0,0)(40,69.28)\ArrowLine(40,69.28)(80,0)\ArrowLine(80,0)(50,0)\ArrowLine(30,0)(0,0)
\DashArrowLine(80,0)(95,-25.98){3}\Photon(40,69.28)(40,94.28){1}{4}\put(47,-2.7){$\times$}
\put(26.8,-2.7){$\times$}\put(41,70){$o$}\put(80.5,0.5){$x$}\put(-5.5,0.5){$y$}

\DashArrowLine(185,-25.98)(200,0){3} \ArrowLine(200,0)(240,69.28)
\ArrowLine(200,0)(240,69.28)\ArrowLine(240,69.28)(255,43.3)
\ArrowLine(265,25.98)(280,0) \ArrowLine(280,0)(200,0)
\DashArrowLine(280,0)(295,-25.98){3}\Photon(240,69.28)(240,94.28){1}{4}
\put(250.2,41.6){$\times$}\put(261,23.8){$\times$}
\put(241,70){$o$}\put(280.5,0.5){$x$}\put(194.5,0.5){$y$}
\put(33,-40){$(a)$}\put(233,-40){$(b)$}
\end{picture}
\end{center}
\vspace{0.5cm}
\caption{\small Diagrams for the contributions of non-local
operators $\bar{\Psi}(x)\Psi(y)$ and $\bar{\Psi}(0)\Psi(x)$.}
\end{figure}
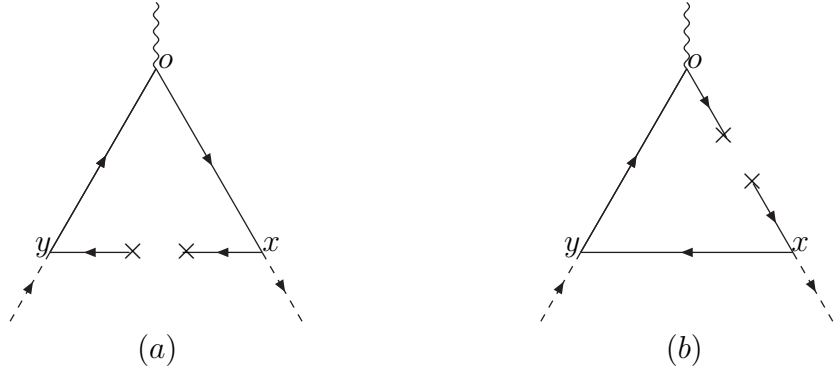
\begin{equation}
\Pi_{\mu\nu}^{2a}=i^2\int d^4 x d^4 y e^{ip_2\cdot x-ip_1\cdot y}
  \langle 0 |\bar{\Psi}(x)\gamma_\nu iS_{F}^s(x)\gamma_\mu
  (1-\gamma_5) iS_F^c(-y)i\gamma_5 \Psi (y)|0\rangle ,
\end{equation}
where $iS_{F}^s(x)$ and $iS_F^c(-y)$ are the propagators of $s$
and $c$ quarks, respectively.  Move the quark field operators
$\bar{\Psi}(x)$ and $\Psi(y)$ together, we get
\begin{equation}
\Pi_{\mu\nu}^{2a}=i^2\int d^4 x d^4 y e^{ip_2\cdot x-ip_1\cdot y}
  \langle 0 |\bar{\Psi}_\alpha (x) \Psi_\beta (y)  |0\rangle
  [\gamma_\nu iS_{F}^s(x)\gamma_\mu (1-\gamma_5) iS_F^c(-y)i\gamma_5]_{\alpha\beta} ,
  \label{pi2a}
\end{equation}
where $\alpha$ and $\beta$ are Dirac spinor indices. The matrix
element $\langle 0 |\bar{\Psi}_\beta (x) \Psi_\alpha (y)
|0\rangle$ can be dealt with in the fixed-point gauge \cite{fix}.
We expand it up to the order of $x^3$ and $y^3$ using the
technique explained in \cite{AAO,ioffe, smilga},
\begin{eqnarray}
&&\langle 0 |\bar{\Psi}^a_\alpha (x)\Psi^b_\beta (y)|0\rangle
 =\delta_{ab}\left[ \langle \bar{\Psi} \Psi\rangle\left(
   \frac{1}{12}\delta_{\beta\alpha}+i\frac{m}{48}
   (\not{x}-\not{y})_{\beta\alpha}-\frac{m^2}{96}
 (x-y)^2\delta_{\beta\alpha}\right.\right. \nonumber\\
   &&\left.\left.-\frac{i}{3!}\frac{m^3}{96}
 (x-y)^2(\not{x}-\not{y})_{\beta\alpha} \right)
 +g\langle \bar{\Psi} TG\sigma\Psi\rangle\left(\frac{1}{192}(x-y)^2
 \delta_{\beta\alpha}\right.\right. \nonumber\\
   &&\left.\left.+\frac{i}{3!}\frac{m}{192}
 (x-y)^2(\not{x}-\not{y})_{\beta\alpha}\right)
 -\frac{i}{3!}\frac{g^2}{3^4\times 2^4}\langle
 \bar{\Psi}\Psi\rangle ^2 (x-y)^2(\not{x}-\not{y})_{\beta\alpha}
 \right. \nonumber \\&& \left.\frac{}{}+\cdots \right],
 \label{qq}
 \end{eqnarray}
$a$ and $b$ in the above are the color indices, $m$ is the quark
mass, and the ellipsis stands for terms of higher orders in $x$
and $y$ expansion. From Eq.~(\ref{qq}) we know that Fig.2(a)
contributes to the coefficients of quark condensate $\langle
\bar{\Psi}\Psi\rangle$, mixed quark-gluon condensate $g\langle
\bar{\Psi} TG\sigma\Psi\rangle$ and the four-quark condensate
$\langle \bar{\Psi}\Psi\rangle ^2$. Substitute Eq.~(\ref{qq}) into
(\ref{pi2a}) and integrate over the coordinates $x$ and $y$, we
can obtain explicitly the coefficients of these condensates
contributed by Fig.2(a).

\subsection*{3.3 Contributions of bi-Gluon Operator
  $G^a_{\mu\nu}G^{a\mu\nu}$}

The diagrams for the contribution of bi-gluon operator are
depicted in Fig.3. They are calculated in the fixed-point gauge,
in which the gauge fixing condition is taken to be
$x^\mu A^a_\mu(x)=0$ \cite{fix}. Then the external gauge field
can be expressed directly in terms of the color field strength
tensor \cite{shif},
\begin{equation}
A^a_\mu (x)=\int^1_0 d\alpha \alpha x^\rho G^a_{\rho\mu}(\alpha
x),
\end{equation}
which is expanded to the first order to be,
\begin{equation}
A^a_\mu (x)=\frac{1}{2}x^\rho  G^a_{\rho\mu}(0)+\cdots.
\end{equation}
In the following calculation, it is convenient to transform
$A^a_\mu (x)$ to the momentum space,
\begin{equation}
A^a_{\mu}(k)=-\frac{i}{2} (2\pi)^2\frac{\partial}{\partial k_\rho}
  \delta ^4(k) G^a_{\rho\mu}(0)+\cdots .
\end{equation}

\begin{figure}[h]
\vspace{1cm}
\begin{center}
\begin{picture}(300,200)(-15,-25.98)
\DashLine(-11.25,100.515)(0,120){3} \Line(0,120)(30,171.96)
\Line(0,120)(30,171.96)\Line(30,171.96)(60,120)\Line(60,120)(0,120)
\DashLine(60,120)(71.25,100.515){3}\Photon(30,171.96)(30,190.71){1}{4}

\GlueArc(0,120)(30,40,60){2}{2}\GlueArc(0,120)(30,0,20){2}{2}
\put(19.88,137,25){$\times$}\put(23.5,129){$\times$}

\DashLine(88.75,100.515)(100,120){3} \Line(100,120)(130,171.96)
\Line(100,120)(130,171.96)\Line(130,171.96)(160,120)\Line(160,120)(100,120)
\DashLine(160,120)(171.25,100.515){3}\Photon(130,171.96)(130,190.71){1}{4}

\GlueArc(130,171.96)(30,240,260){2}{2}\GlueArc(130,171.96)(30,280,300){2}{2}
\put(120.4,138.8){$\times$}\put(131.3,138.8){$\times$}

\DashLine(188.75,100.515)(200,120){3} \Line(200,120)(230,171.96)
\Line(200,120)(230,171.96)\Line(230,171.96)(260,120)\Line(260,120)(200,120)
\DashLine(260,120)(271.25,100.515){3}\Photon(230,171.96)(230,190.71){1}{4}

\GlueArc(260,120)(30,160,180){2}{2}\GlueArc(260,120)(30,120,140){2}{2}
\put(232.3,136){$\times$}\put(228.125,128){$\times$}

\DashLine(-11.25,-19.485)(0,0){3} \Line(0,0)(30,51.96)
\Line(0,0)(30,51.96)\Line(30,51.96)(60,0)\Line(60,0)(0,0)
\DashLine(60,0)(71.25,-19.485){3}\Photon(30,51.96)(30,70.71){1}{4}

\Gluon(11.25,19.485)(-1.74,26.985){2}{2}\Gluon(18.75,32.475)(5.76,39.975){2}{2}

\put(-5.625,23.5){$\times$}\put(1.65,36.5){$\times$}
 \DashLine(88.75,-19.485)(100,0){3} \Line(100,0)(130,51.96)
\Line(100,0)(130,51.96)\Line(130,51.96)(160,0)\Line(160,0)(100,0)
\DashLine(160,0)(171.25,-19.485){3}\Photon(130,51.96)(130,70.71){1}{4}

\Gluon(122.5,0)(122.5,-15){2}{2}\Gluon(137.5,0)(137.5,-15){2}{2}

\put(119.875,-16.5){$\times$}\put(135.25,-16.5){$\times$}

\DashLine(188.75,-19.485)(200,0){3} \Line(200,0)(230,51.96)
\Line(200,0)(230,51.96)\Line(230,51.96)(260,0)\Line(260,0)(200,0)
\DashLine(260,0)(271.25,-19.485){3}\Photon(230,51.96)(230,70.71){1}{4}

\Gluon(248.75,19.485)(261.74,26.985){2}{2}\Gluon(241.25,32.475)(254.24,39.975){2}{2}

\put(256.5,24.75){$\times$}\put(250,37.5){$\times$}

\put(23,90){(a)}
 \put(125,90){(b)}
 \put(225,90){(c)}

 \put(23,-35){(d)}
 \put(125,-35){(e)}
 \put(225,-35){(f)}
\end{picture}
\end{center}
\vspace{0.5cm}
\caption{\small Diagrams for contributions of bi-gluon operator.}
\end{figure}
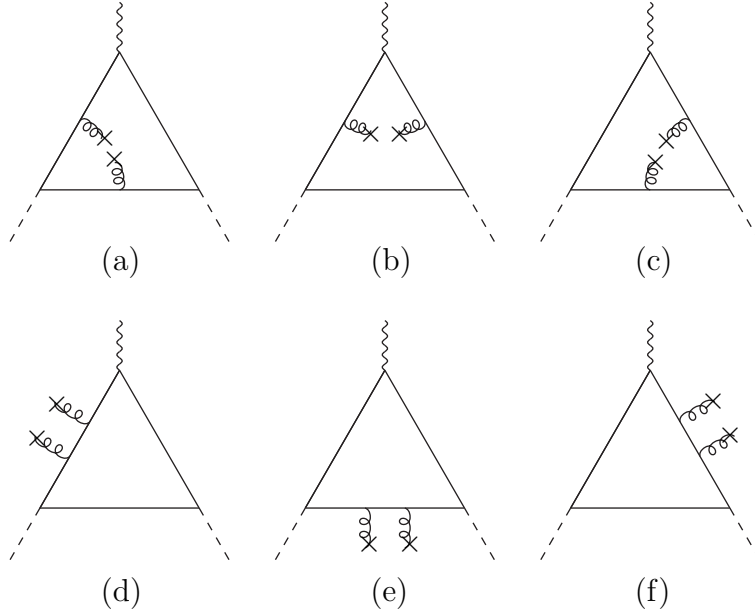

Then the amplitude can be written down in the momentum space
by following the standard Feynman rule. Again, as what we
did in the previous subsections, we move the gluon strength
tensor operator together: $G^a_{\alpha\sigma}G^b_{\beta\rho}$.
Then using the following decomposition to obtain the bi-gluon
condensate,
\begin{equation}
\langle 0|G^a_{\alpha\sigma}G^b_{\beta\rho}|o\rangle=\frac{1}{96}
 \langle GG \rangle \delta_{ab}(g_{\alpha\beta}g_{\sigma\rho}-
  g_{\alpha\rho}g_{\sigma\beta}),
\end{equation}
in which $\langle GG \rangle$ is the abbreviation of
$\langle 0|G^a_{\mu\nu}G^{a\mu\nu}|0\rangle$.

In the evaluation of the diagrams of Fig.3 some types of loop
integrals encountered are treated at first by derivatives with
respect to the quark masses, then transform them to dispersion
integrals by using Cutkosky's rule and with help of $I$, $I_\mu$ and
$I_{\mu\nu}$ functions given previously. For instance,
\begin{eqnarray}
&&\int d^4 k \frac{k_\mu
  k_\nu}{(k^2-m^2)[(k+p_2)^2-m_2^2]^2[(k+p_1)^2-m_1^2]^2}\nonumber\\
&=&\frac{\partial}{\partial m_2^2}\frac{\partial}{\partial m_1^2}
  \int d^4 k \frac{k_\mu
  k_\nu}{(k^2-m^2)[(k+p_2)^2-m_2^2][(k+p_1)^2-m_1^2]}\nonumber\\
&=&-2\pi i\frac{\partial}{\partial m_2^2}\frac{\partial}{\partial m_1^2}
  \int ds_1 ds_2 \frac{I_{\mu\nu}}{(s_1-p_1^2)(s_2-p_2^2)},
\end{eqnarray}
and
\begin{eqnarray}
&&\int d^4 k \frac{k_\mu
  k_\nu k\cdot p_2}{(k^2-m^2)^2[(k+p_2)^2-m_2^2]^2[(k+p_1)^2-m_1^2]^2}\nonumber\\
&=&\frac{\partial}{\partial m_2^2}\frac{\partial}{\partial m_1^2}
  \int d^4 k \frac{k_\mu
  k_\nu k\cdot p_2}{(k^2-m^2)^2[(k+p_2)^2-m_2^2][(k+p_1)^2-m_1^2]}\nonumber\\
&=&-2\pi i\frac{\partial}{\partial m^2}\frac{\partial}
  {\partial m_2^2}\frac{\partial}{\partial m_1^2}
  \int ds_1 ds_2 \frac{-\frac{1}{2}(s_2-m_2^2+m^2)I_{\mu\nu}}{(s_1-p_1^2)(s_2-p_2^2)},
\end{eqnarray}
where the term $-\frac{1}{2}(s_2-m_2^2+m^2)$ comes from the
$\delta$ functions $\delta (k^2-m^2)\delta[(k+p_2)^2-m_2^2]$ with
the substitution $p_2^2\to s_2$ when using the Cutkosky's rule.

After some long steps of calculation, we finally find that the
contributions of the diagrams (a) to (f) in Fig.3 cancel each
other. Therefore there are no contributions of gluon condensate in
$D_s\to \Phi$ transition.

\subsection*{3.4 Contributions of Quark-Gluon mixing and
Four-Quark Operators: $\bar{\Psi}(x) \Psi (y)G^a_{\mu\nu}$ and
$\langle \bar{\Psi}\Psi\rangle ^2$}

The diagrams for quark-gluon mixing and four-quark contributions
are depicted in Fig.4 and Fig.5, respectively. The techniques are
similar to that explained in previous subsections. We only give
some different points here.
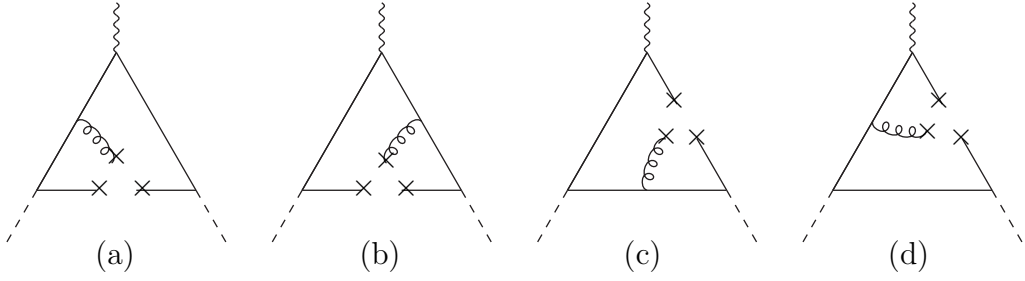
\begin{figure}[h]
\vspace{1cm}
\begin{center}
\begin{picture}(400,100)(-15,-25.98)
\DashLine(-11.25,-19.485)(0,0){3} \Line(0,0)(30,51.96)
\Line(0,0)(30,51.96)\Line(30,51.96)(60,0)\Line(60,0)(37.5,0)\Line(22.5,0)(0,0)
\DashLine(60,0)(71.25,-19.485){3}\Photon(30,51.96)(30,70.71){1}{4}\put(35.25,-2.025){$\times$}
\put(19,-2.025){$\times$}
\GlueArc(0,0)(30,20,60){2}{3}\put(25.5,10){$\times$}

\DashLine(88.75,-19.485)(100,0){3} \Line(100,0)(130,51.96)
\Line(100,0)(130,51.96)\Line(130,51.96)(160,0)\Line(160,0)(137.5,0)\Line(122.5,0)(100,0)
\DashLine(160,0)(171.25,-19.485){3}\Photon(130,51.96)(130,70.71){1}{4}\put(135,-2.025){$\times$}
\put(119,-2.025){$\times$}
\GlueArc(160,0)(30,120,160){2}{3}\put(127.5,8.5){$\times$}
\DashLine(188.75,-19.485)(200,0){3} \Line(200,0)(230,51.96)
\Line(200,0)(230,51.96)\Line(230,51.96)(241.25,32.475)
\Line(248.75,19.485)(260,0) \Line(260,0)(200,0)
\DashLine(260,0)(271.25,-19.485){3}\Photon(230,51.96)(230,70.71){1}{4}
\put(236.5,31.2){$\times$}\put(245,17){$\times$}

\GlueArc(260,0)(30,140,180){2}{3}\put(233.5,17.5){$\times$}
\DashLine(288.75,-19.485)(300,0){3} \Line(300,0)(330,51.96)
\Line(300,0)(330,51.96)\Line(330,51.96)(341.25,32.475)
\Line(348.75,19.485)(360,0) \Line(360,0)(300,0)
\DashLine(360,0)(371.25,-19.485){3}\Photon(330,51.96)(330,70.71){1}{4}

\put(336.5,31.2){$\times$}\put(345,17){$\times$}

\GlueArc(330,51.96)(30,240,280){2}{3}\put(332.25,19.5){$\times$}
\put(22,-28){(a)}
 \put(122,-28){(b)}
 \put(222,-28){(c)}

\put(322,-28){(d)}
\end{picture}
\end{center}
\caption{Diagrams for mixed  quark-gluon operators.}
\end{figure}

\begin{figure}[h]
\vspace{1cm}
\begin{center}
\begin{picture}(300,200)(-15,-25.98)
\DashLine(-11.25,100.515)(0,120){3} \Line(0,120)(30,171.96)
\Line(0,120)(30,171.96)\Line(60,120)(37.5,120)\Line(22.5,120)(0,120)
\DashLine(60,120)(71.25,100.515){3}\Photon(30,171.96)(30,190.71){1}{4}
\put(33.3,117.975){$\times$}
\put(17.5,117.975){$\times$} \Line(30,171.96)(41.25,152.475)
\Line(48.75,139.485)(60,120)
\put(36.5,151.2){$\times$}\put(45,137){$\times$}

\Gluon(15,147.6)(48,120){1.6}{8}
 \DashLine(88.75,100.515)(100,120){3} \Line(100,120)(130,171.96)
\Line(100,120)(130,171.96)\Line(160,120)(137.5,120)\Line(122.5,120)(100,120)
\DashLine(160,120)(171.25,100.515){3}\Photon(130,171.96)(130,190.71){1}{4}
\put(133.3,117.975){$\times$} \put(117.5,117.975){$\times$}
\Line(130,171.96)(141.25,152.475) \Line(148.75,139.485)(160,120)
\put(136.5,151.2){$\times$}\put(145,137){$\times$}

\Gluon(115,147.6)(154.375,128.8){1.6}{8}
\DashLine(188.75,100.515)(200,120){3} \Line(200,120)(230,171.96)
\Line(200,120)(230,171.96)\Line(260,120)(237.5,120)\Line(222.5,120)(200,120)
\DashLine(260,120)(271.25,100.515){3}\Photon(230,171.96)(230,190.71){1}{4}
\put(233.3,117.975){$\times$}
\put(217.5,117.975){$\times$} \Line(230,171.96)(241.25,152.475)
\Line(248.75,139.485)(260,120)
\put(236.5,151.2){$\times$}\put(245,137){$\times$}

\GlueArc(230,120)(15,0,180){1.6}{8} \DashLine(-11.25,-19.485)(0,0){3}
\Line(0,0)(30,51.96)
\Line(0,0)(30,51.96)\Line(60,0)(37.5,0)\Line(22.5,0)(0,0)
\DashLine(60,0)(71.25,-19.485){3}\Photon(30,51.96)(30,70.71){1}{4}
\put(33.3,-2.5){$\times$}
\put(18,-2.5){$\times$} \Line(30,51.96)(41.25,32.475)
\Line(48.75,19.485)(60,0)
\put(36.5,31.2){$\times$}\put(45,17){$\times$}
\Gluon(11.25,0)(54.375,9.675){1.6}{8}
\DashLine(88.75,-19.485)(100,0){3} \Line(100,0)(130,51.96)
\Line(100,0)(130,51.96)\Line(160,0)(137.5,0)\Line(122.5,0)(100,0)
\DashLine(160,0)(171.25,-19.485){3}\Photon(130,51.96)(130,70.71){1}{4}
\put(133.3,-2.5){$\times$}
\put(117.5,-2.5){$\times$} \Line(130,51.96)(141.25,32.475)
\Line(148.75,19.485)(160,0)
\put(136.5,31.2){$\times$}\put(145,17){$\times$}

\Gluon(134.125,42.75)(145,0){1.6}{8}
\DashLine(188.75,-19.485)(200,0){3} \Line(200,0)(230,51.96)
\Line(200,0)(230,51.96)\Line(260,0)(237.5,0)\Line(222.5,0)(200,0)
\DashLine(260,0)(271.25,-19.485){3}\Photon(230,51.96)(230,70.71){1}{4}
\put(233.3,-2.5){$\times$}
\put(217.5,-2.5){$\times$} \Line(230,51.96)(241.25,32.475)
\Line(248.75,19.485)(260,0)
\put(236.5,31.2){$\times$}\put(245,17){$\times$}

\GlueArc(245,29.7)(15,126,293){1.6}{8}

\put(23,90){(a)}
 \put(125,90){(b)}
 \put(225,90){(c)}

 \put(23,-28){(d)}
 \put(125,-28){(e)}
 \put(225,-28){(f)}
\end{picture}
\end{center}
\caption{\small diagrams for four-quark contributions.}
\end{figure}
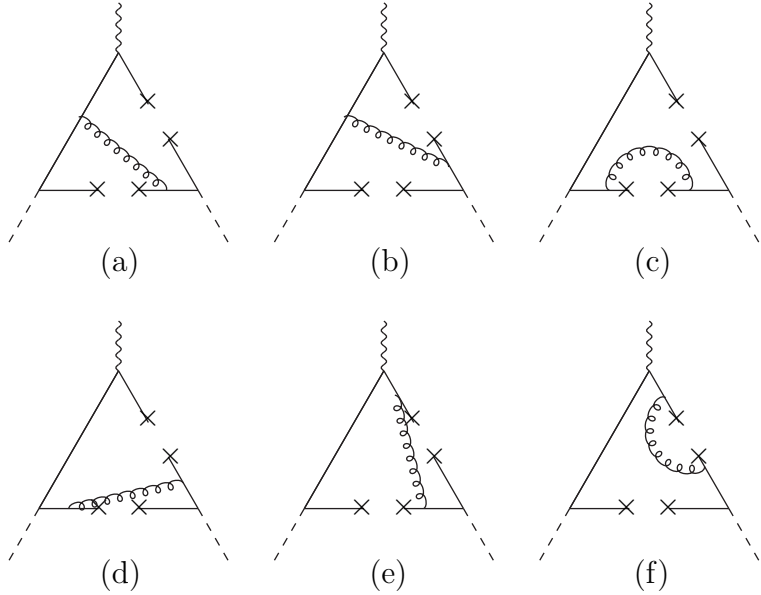

The vacuum average of non-local quark-gluon mixing operator
$\bar{\Psi}(x) \Psi (y)G^a_{\mu\nu}$ is calculated to be
\begin{eqnarray}
&&\langle 0|\bar{\Psi}^i_\alpha (x) \Psi ^j_\beta (y)
G^a_{\mu\nu}|0\rangle \nonumber\\
&=&\frac{1}{192}\langle \bar{\Psi}\sigma T G\Psi \rangle
(\sigma_{\mu\nu})_{\beta\alpha}T^a_{ji} + \left[-\frac{g}{96\times
9}\langle\bar{\Psi}\Psi\rangle^2
(g_{\rho\mu}\gamma_\nu-g_{\rho\nu}\gamma_\mu)(x+y)^\rho\right.\nonumber\\
&&\left.+i(y-x)^\rho\left(\frac{g}{96\times
9}\langle\bar{\Psi}\Psi\rangle^2 +\frac{m}{96\times 4}\langle
\bar{\Psi}\sigma T G\Psi \rangle
\right)\varepsilon_{\rho\mu\nu\sigma}\gamma_5\gamma^\sigma
\right]_{\beta\alpha}T^a_{ji},
\end{eqnarray}
where $\langle \bar{\Psi}\sigma TG\Psi \rangle$ and
$\langle\bar{\Psi}\Psi\rangle^2$ are the abbreviations of $\langle
0| \bar{\Psi}\sigma_{\mu\nu} T^aG^{a\mu\nu}\Psi |0 \rangle$ and
$\langle 0|\bar{\Psi}\Psi |0\rangle^2$ respectively. $g$ is the
strong coupling.

Because we calculate up to the condensate of dimension-six
operators, the external gluon field $A^a_{\mu}(x)$ in Fig.4
should be expanded up to the second term, which will contribute a
dimension-six operator,
\begin{eqnarray}
A^a_{\mu}(x)&=&\int^1_0d\alpha \alpha x^\rho
G^a_{\rho\mu}(\alpha x)\nonumber\\
&=&\frac{1}{2}x^\rho G^a_{\rho\mu}(0)+\frac{1}{3}x^\alpha x^\rho
\hat{D}_\alpha G^a_{\rho\mu}(0)+\cdots ~,
\end{eqnarray}
where $\hat{D}_\alpha$ is the covariant derivative in the adjoint
representation, $(\hat{D}_\alpha)^{mn} =\partial_\alpha
\delta^{mn}-gf^{amn}A^a_\alpha$. Then another vacuum matrix
element needed is \cite{ioffe}
\begin{equation}
\langle 0|\bar{\Psi}^i_\alpha \Psi^j_\beta \hat{D}_\xi
G^a_{\sigma\rho} |0\rangle=-\frac{g}{3^3\times 2^4}
\langle \bar{\Psi}\Psi \rangle ^2
(g_{\xi\rho}\gamma_\sigma-g_{\xi\sigma}\gamma_\rho )_{\beta\alpha}
T^a_{ji}~.
\end{equation}
We calculate these diagrams and find that the contributions of
Fig.4(c), (d) and Fig.5(c), (d) vanish after double Borel
transformation in two variables $p_1^2$ and $p_2^2$, because only
one variable appearing in the denominator, for instance,
$\frac{1}{q^2 (p_2^2-m_2^2)}$. The Borel transformation in $p_1^2$
will kill such terms.

\vspace{1cm}

Following the above method, after some tedious algebraic
derivation with the software MATHEMATICA, we obtain the
coefficients $f_0$, $f_1+f_3$, $f_1-f_3$ and $f_5$ needed in
Eq.~(\ref{formsr}). They are listed in Appendix.

\section*{4. Numerical Analysis and Discussion}

In the numerical analysis the standard values of the
condensates at the renormalization point $\mu =1\mbox{GeV}$ are
taken \cite{SVZ, Narison},
\begin{eqnarray}
&\langle \bar{q}q\rangle =-(0.24\pm 0.01 \mbox{GeV})^3, ~~~~
\langle \bar{s}s\rangle =m_0^2 \langle \bar{q}q\rangle,
\nonumber\\[4mm]
 & g\langle \bar{\Psi}\sigma T\Psi \rangle =m_0^2 \langle
\bar{\Psi}\Psi \rangle, ~~~~\alpha_s\langle \bar{\Psi}\Psi\rangle
^2= 6.0\times10^{-5}\mbox{GeV}^6 ,\\[4mm]
& m_0^2=0.8\pm 0.2 \mbox{GeV}^2. \nonumber
\end{eqnarray}
The quark masses are fixed to be $m_s=140\mbox{MeV}$,
$m_c=1.3\mbox{GeV}$ \cite{175}, and the decay constant of
$\phi$ meson is extracted from experimental data $f_{\phi}=0.228$ \cite{PDG}.
For the decay constant of $D_s$ meson we take $f_{D_s}=0.214\pm
0.038\mbox{GeV}$ \cite{175}.

\begin{figure}
 \begin{center}
 \epsfig{file=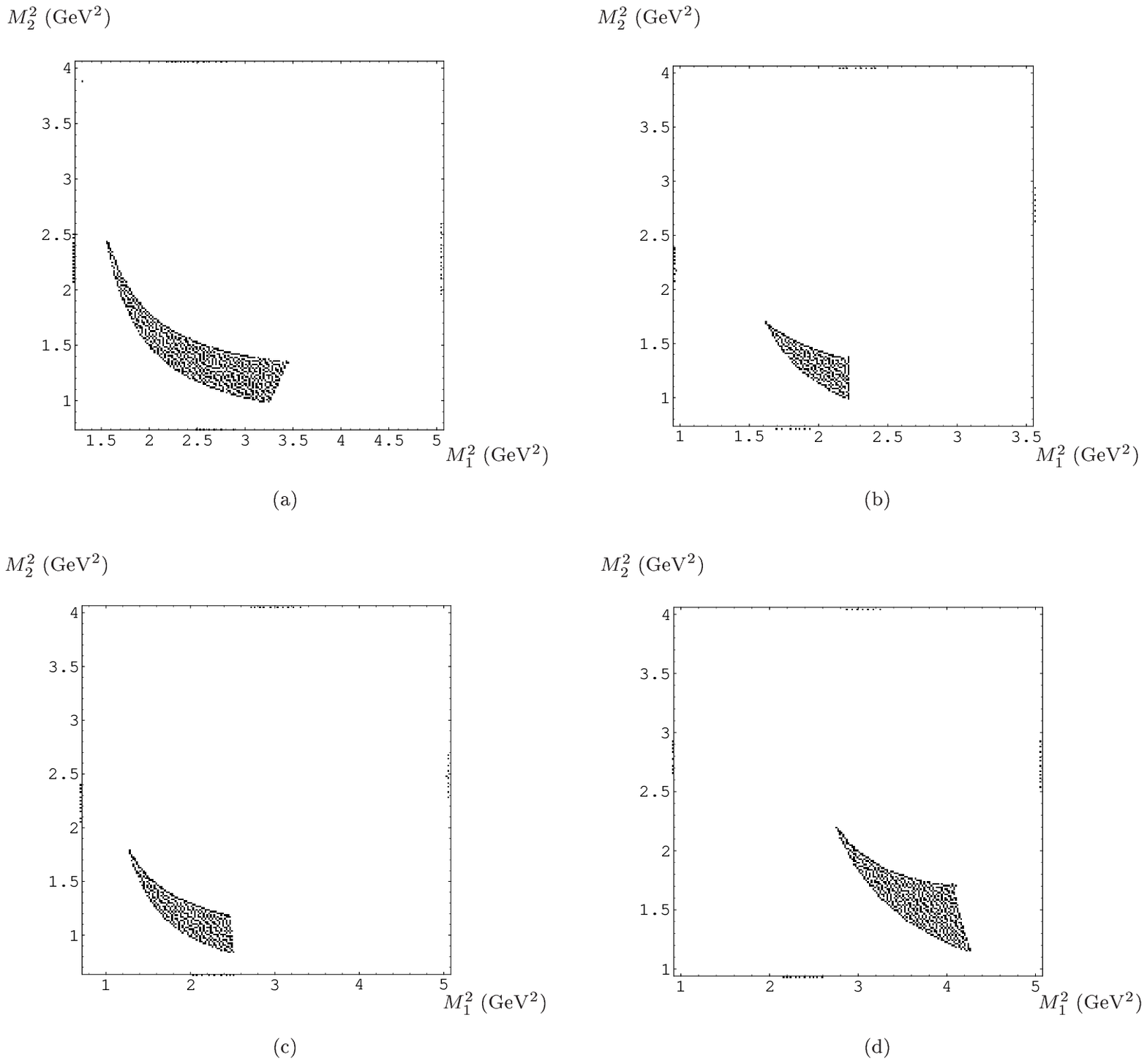, width=14cm,height=14cm}
  \end{center}
\caption{\small Selected regions of $M_1^2$ and $M_2^2$: (a) for
$V$; (b) for $A_0$; (c) for $A_1$; (d) for $A_2$.} \label{region}
\end{figure}

The Borel parameters $M_1$ and $M_2$ are not physical parameters.
The physical result should not depend on them if the operator
product expansion can be calculated up to infinite order. However,
OPE has to be truncated to some finite orders in practice.
Therefore, Borel parameters have to be selected in some ``windows"
to get the best stability of the physical results. The requirement
to select the stable ``windows" is:  the Borel parameters can not
be too large, or, contributions of higher resonance and continuum
states can not be effectively suppressed; at the same time, they
should not be too small, or, the truncated OPE would fail because
the series in OPE generally depend on Borel parameters in the
denominator $1/M$. We find the optimal stability with the
requirements shown in Table \ref{T1} and the thresholds $s_1^0$,
$s_2^0$ in the ranges $s_1^0=5.8-6.2 \mbox{GeV}^2$, $s_2^0=1.9-2.1
\mbox{GeV}^2$. The regions of Borel parameters which satisfies the
requirements of Table \ref{T1} are shown in Fig.\ref{region} in
two-dimensional diagram of $M_1^2$ and $M_2^2$. We find good
stability of the form factors within these regions.
\begin{tiny}
\begin{table}[h]
\caption{Requirements to select Borel Parameters $M_1^2$ and $M_2^2$
 for each form factors $V(0)$, $A_0(0)$, $A_1(0)$ and $A_2(0)$}
\begin{center}
\begin{tabular}{|c|c|c|c|}\hline
Form Factors & contribution  & continuum of & continuum of  \\
             &of condensate &  $D_s$ channel & $\phi$ channel\\
             \hline
$V(0)$  & $\le 49\% $ & $\le 5\%$ & $\le 26\%$ \\ \hline $A_0(0)$&
$\le 29\% $ & $\le 16\%$ & $\le 31\%$ \\ \hline $A_1(0)$ &$\le
49\% $ & $\le 18\%$ & $\le 22\%$ \\ \hline $A_2(0)$  & $\le 11\% $
& $\le 27\%$ & $\le 5\%$ \\ \hline
\end{tabular}\end{center}
\label{T1}
\end{table}
\end{tiny}

Because it is not easy to show the contribution of each term of
OPE in two-dimensional regions of $M_1^2$ and $M_2^2$, we show the
contributions of  perturbative and condensate terms in Table
\ref{T2} at a representative point $(M_1^2,~M_2^2)$ in the stable
region of $M_1^2$ and $M_2^2$.  In general the higher the
dimension of the operators, the smaller the relevant contributions
of the condensates. The main contributions to $V(0)$, $A_1(0)$ and
$A_2(0)$ are from perturbative and quark condensate term. For
$A_0(0)$, the largest two contributions are from perturbative term
and mixed quark-gluon condensate. Contributions of four-quark
condensate are less than a few percent, therefore contributions of
operator of dimension 6 are negligible.

\begin{tiny}
\begin{table}[h]
\caption{Contributions of perturbative and condensate terms in the
operator-product expansion to the form factors $V(0)$, $A_0(0)$,
$A_1(0)$ and $A_2(0)$, at a representative point $(M_1^2,~M_2^2)$
in the stable region of $M_1^2$ and $M_2^2$. $f^{pert}$:
perturbative; $f^{(3)}$: quark condensate; $f^{(4)}$: gluon
condensate; $f^{(5)}$: mixed quark-gluon condensate; $f^{(6)}$:
four-quark condensate.}
\begin{center}
\begin{tabular}{|c|c|c|c|c|c|c|c|}\hline
Form Factors &total &$f^{pert}$ &$f^{(3)}$ &$f^{(4)}$ &$f^{(5)}$ &$f^{(6)}$ &$(M_1^2,M_2^2)~\mbox{GeV}^2$ \\
                         \hline
$V(0)$    & $1.20 $ & $0.63$ & $0.66$ & $0$ & $-0.10$ &$0.01$ &
$(2.2,1.4)$ \\ \hline $A_0(0)$  & $0.43 $ & $0.28$ & $-0.10$ & $0$
& $0.23$ &$0.02$ & $(1.7,1.5)$ \\ \hline
 $A_1(0)$  & $0.53 $&$0.28$ & $0.20$ & $0$ & $0.04$  &$0.01$ & $(2.0,1.2)$ \\ \hline
$A_2(0)$  & $0.57 $ & $0.22$ & $0.44$ & $0$ & $-0.09$ &$0.00$ &
$(3.6,1.5)$\\ \hline
\end{tabular}\end{center}
\label{T2}
\end{table}
\end{tiny}

The final results for the form factors at $q^2=0$ are
\begin{eqnarray}
V(0)&=& 1.21\pm 0.33,~~~~~~~~A_0(0)~=~0.42\pm 0.12,\nonumber
\\[4mm]
A_1(0)&=& 0.55\pm 0.15,~~~~~~~A_2(0)~=~0.59\pm 0.17, \\[4mm]
r_V\equiv \frac{V(0)}{A_1(0)}&=& 2.20\pm 0.85,~~~~~~r_2\equiv
\frac{A_2(0)}{A_1(0)}~=~1.07\pm 0.43 .\nonumber
\end{eqnarray}

We compare our results for the ratios of form factors with
experimental data in Table \ref{T3}. It shows that the results are consistent
with experimental data.
\begin{tiny}
\begin{table}[h]
\caption{Comparison of our results of $r_V$ and $r_2$ with experimental
data: E791 is from Ref.\cite{exp4}, CLEO from Ref.\cite{exp3}, E687 from
Ref.\cite{exp2} and E653 from Ref.\cite{exp1}.}
\begin{center}
\begin{tabular}{|c|c|c|}  \hline
      & $r_V$                  & $r_2$ \\ \hline
E791  & $2.27\pm 0.35\pm 0.22$ & $1.57\pm 0.25\pm 0.19$\\ \hline
CLEO  & $0.9\pm 0.6\pm 0.3$    & $1.4\pm 0.5\pm 0.3$\\  \hline
E687  & $1.8\pm 0.9\pm 0.2$    & $1.1\pm 0.8\pm 0.1$\\  \hline
E653  & $2.3^{+1.1}_{-0.9}\pm 0.4$ & $2.1^{+0.6}_{-0.5}\pm 0.2$\\
\hline Average &  $1.92\pm 0.32$  &  $1.60\pm 0.24$\\ \hline our
result &  $2.20\pm 0.85 $ & $1.07\pm 0.43 $\\ \hline
\end{tabular}\end{center}
\label{T3}
\end{table}
\end{tiny}

\begin{figure}[h]
 \begin{center}
 \epsfig{file=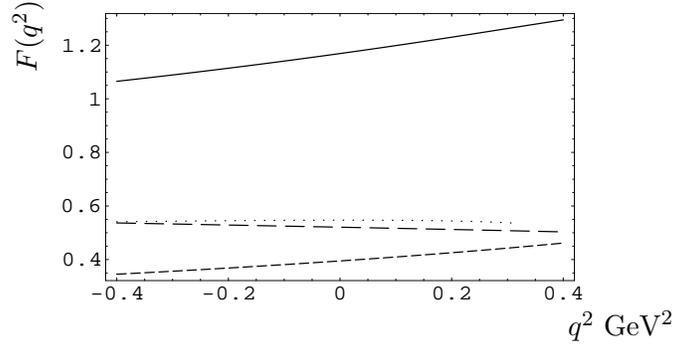, width=7cm,height=4cm}
 \begin{picture}(0,30)
 \put(-10,-10){\small{$q^2~\mbox{GeV}^2$}}
 \put(-220,90){\rotatebox{90}{\small{$F(q^2)$} }}
 \end{picture}
 \end{center}
\caption{\small $q^2$ dependence of the form factors from QCD sum rule.
The solid curve is for $V(q^2)$, the short dashed curve for
$A_0(q^2)$, the long dashed curve for $A_1(q^2)$, and the dotted
one is for $A_2(q^2)$.} \label{q2}
\end{figure}

The physical region for $q^2$ in $D_s\to \phi\bar{\ell}\nu$ decay
extends from $0$ to $(m_{D_s}-m_\phi )^2\simeq 0.9~\mbox{GeV}^2$.
In the range $q^2<0.4 \mbox{GeV}^2$, there is no non-Landau-type
singularity \cite{PBD} with the thresholds $s_1^0$ and $s_2^0$
chosen in this paper. The $q^2$ dependence of the form factors is
shown in Fig.\ref{q2} in the range $-0.4\mbox{GeV}^2
<q^2<0.4\mbox{GeV}^2$. Within this range, the behavior of $V(q^2)$
and $A_0(q^2)$ is well compatible with the pole-model,
$$ V(q^2)=\frac{V(0)}{1-q^2/m_{pole}^V}. $$
While the $q^2$ dependence of and $A_1(q^2)$ and $A_2(q^2)$ is
very weak.

We fit $V(q^2)$ and $A_0(q^2)$ by the pole model in the range
$-0.4\mbox{GeV}^2 <q^2<0.4\mbox{GeV}^2$, and extrapolate the
fitted result to the whole physical region. The fitted pole masses
are,
\begin{eqnarray}
m^V_{pole}~=& 2.08\pm 0.13~\mbox{GeV},\nonumber\\
m^{A_0}_{pole}~=& 1.9\pm 0.2~\mbox{GeV}.
\end{eqnarray}

The form factors calculated in QCD sum rule in this paper are used
to calculate the differential and total decay rate of $D_s\to
\phi\bar{\ell}\nu$ decay. There are three polarization states for
$\phi$ meson: one longitudinal state, two transverse polarization
states (right-handed and left-handed). The differential decay rate
to longitudinally polarized $\phi$ meson is
\begin{eqnarray}
\frac{d\Gamma_L}{d
q^2}~=&\displaystyle\frac{G_F^2|V_{cs}|^2}{192\pi ^3
   m_{D_s}^3} \sqrt{\lambda
   (m_{D_s}^2,m_{\phi}^2,q^2)}\left|\frac{1}{2m_{\phi}}
   \left[(m_{D_s}^2-m_{\phi}^2-q^2)(m_{D_s}+m_{\phi})A_1(q^2)\right.\right.
   \nonumber\\[4mm]
    &\left.\left.-\displaystyle\frac{\lambda (m_{D_s}^2,m_{\phi}^2,q^2)}{m_{D_s}+m_{\phi}}
    A_2(q^2)\right]\right|^2,
\end{eqnarray}
where $G_F$ is Fermi constant, $V_{cs}$ is CKM matrix element for
$c\to s $ transition, and
$$\lambda
(m_{D_s}^2,m_{\phi}^2,q^2)\equiv
(m_{D_s}^2+m_{\phi}^2-q^2)^2-4m_{D_s}^2m_{\phi}^2.
$$
The differential decay rate to transverse state is
\begin{equation}
\frac{d\Gamma_T^\pm}{d
q^2}=\displaystyle\frac{G_F^2|V_{cs}|^2}{192\pi ^3
   m_{D_s}^3} \lambda
   (m_{D_s}^2,m_{\phi}^2,q^2)\left|\frac{V(q^2)}{m_{D_s}+m_{\phi}}
   \mp \frac{(m_{D_s}+m_{\phi})A_1(q^2)}{\sqrt{\lambda
   (m_{D_s}^2,m_{\phi}^2,q^2)}}\right|^2,
\end{equation}
where the symbol $``+"$ and $``-"$ denote right and left-handed
states, respectively. Finally, the combined transverse and total
differential decay rates are
\begin{equation}
\frac{d\Gamma_T}{d q^2}=\frac{d}{dq^2}(\Gamma_T^+ +\Gamma_T^-),
~~~~\frac{d\Gamma}{d q^2}=\frac{d}{dq^2}(\Gamma_L +\Gamma_T).
\end{equation}

\begin{figure}[h]
 \begin{center}
 \epsfig{file=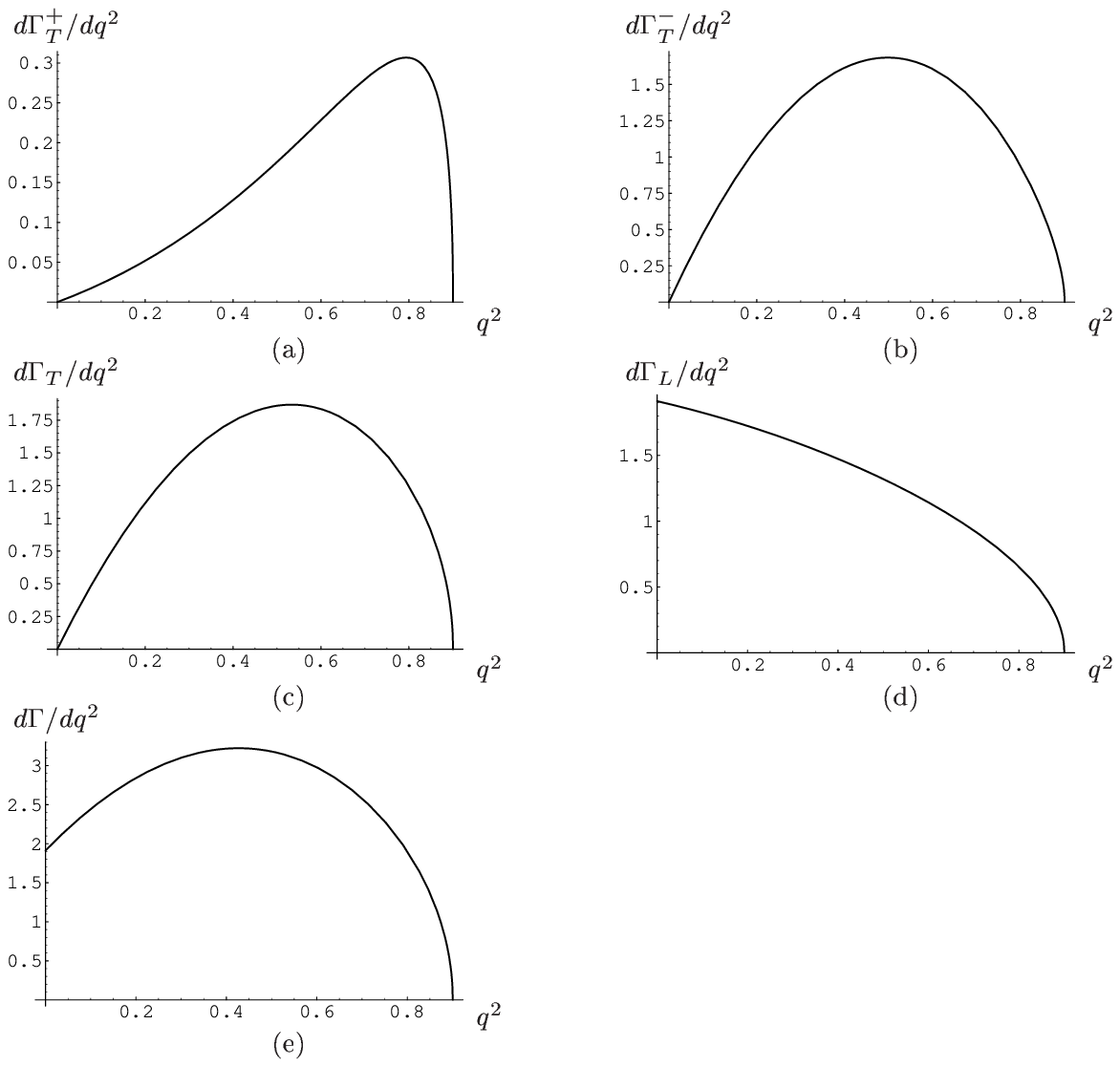, width=12cm,height=12cm}
 \end{center}
\caption{\small Differential decay widths of $D_s^+\to \phi \bar{\ell}\nu$ as a function
of momentum transfer squared $q^2$ in unit of $10^{-14}\mbox{GeV}^{-1}$.} \label{gamma}
\end{figure}

The differential decay widths as a function of momentum transfer
squared $q^2$ are shown in Fig.\ref{gamma}. Integrate them over
$q^2$ in the whole physical region from $q^2=0$ to
$(m_{D_s}-m_{\phi})^2$, we get the integrated decay widths
\begin{eqnarray}
&\Gamma_T^+=(1.39\pm 0.75)\times 10^{-15}\mbox{GeV},~~~~~~
\Gamma_T^-=(1.05\pm 0.22)\times
10^{-14}\mbox{GeV},\nonumber\\[3mm]
&\Gamma_L=(1.18\pm 0.43)\times 10^{-14}\mbox{GeV},~~~~~~
\Gamma_T=(1.19\pm 0.29)\times 10^{-14}\mbox{GeV},
\end{eqnarray}
and the ratio of $\Gamma_L/\Gamma_T$ is
\begin{equation}
\Gamma_L/\Gamma_T =0.99\pm 0.43\; ,
\end{equation}
which is consistent with the averaged experimental data
$(\Gamma_L/\Gamma_T)^{exp}=0.72\pm 0.18 $ \cite{PDG}. The detailed
comparison of this ratio with experimental data is shown in Table
\ref{T4}.

\begin{tiny}
\begin{table}[h]
\caption{\small Comparison of our results of $\Gamma_L/\Gamma_T$ with experimental
data: CLEO from Ref.\cite{exp3}, E687 from
Ref.\cite{exp2} and E653 from Ref.\cite{exp1}.}
\begin{center}
\begin{tabular}{|c|c|}  \hline
      & $\Gamma_L/\Gamma_T$  \\ \hline
CLEO  & $1.0\pm 0.3\pm 0.2$ \\  \hline E687  & $1.0\pm 0.5\pm 0.1$
\\  \hline E653  & $0.54\pm 0.21\pm 0.10$ \\ \hline Average &
$0.72\pm 0.18$  \\ \hline our result &  $0.99\pm 0.43 $ \\ \hline
\end{tabular}\end{center}
\label{T4}
\end{table}
\end{tiny}

We use the total decay width of $D_s$ meson
$\Gamma_{D_s}=1.34\times 10^{-12}$ \cite{PDG} to obtain the
branching ratio of  $D_s^+\to \phi \bar{\ell}\nu$, our result is
\begin{equation}
Br(D_s^+\to \phi \bar{\ell}\nu)=(1.8\pm 0.5)\%,
\end{equation}
which is in good agreement with experimental data $Br(D_s^+\to \phi
\bar{\ell}\nu)^{exp}=(2.0\pm 0.5)\% $.

\section*{5. Summary}

We calculate the transition form factors for $D_s\to \phi$
transition in the region $q^2\le 0.4\mbox{GeV}^2$ in QCD sum rule,
where no non-Landau-type singularity occurs. Then fit the result
from QCD sum rule in this region of momentum transfer, and
extrapolate it to the whole physical region in the decay $D_s^+\to
\phi \bar{\ell}\nu $. We treat the two Borel parameters $M_1^2$
and $M_2^2$ as independent parameters, and select the allowed
region for $M_1^2$ and $M_2^2$ by requiring that the higher
resonance and continuum contributions in $D_s$ and $\phi$ channels
are not large, at the same time requiring that the condensate of
higher dimension operators do not contribute too much. We find
good stability for the transition form factors $V$, $A_0$, $A_1$
and $A_2$ in the relevant two-dimensional regions of $M_1^2$ and
$M_2^2$. We obtain the results of the transition form factors $V$,
$A_0$, $A_1$ and $A_2$ in these regions of $M_1^2$ and $M_2^2$.
Our result of the ratios of these form factors $r_V$ and $r_2$ are
well consistent with experimental data.

We studied the process $D_s^+\to \phi \bar{\ell}\nu $ with the
form factors calculated from QCD sum rule. For the transverse
polarization state of the final $\phi$ meson, the rate of $D_s$
decaying to right-hand state is almost an order smaller than
decaying to left-hand state. The ratio of $\Gamma_L/\Gamma_T$ and
the branching ratio of $D_s^+\to \phi \bar{\ell}\nu $ are in good
agreement with the experimental data within the error bars of both
the present experimental data and theoretical calculation.

\vspace{1cm}

{\bf Acknowledgements} This work is supported in part by the
National Science Foundation of China under contract No.10205017,
90103011, and by the Grant of BEPC National Laboratory.

 \vspace{1cm}

\begin{center}{\bf Appendix }\end{center}

 Borel transformed Coefficients of perturbative and nonperturbative contributions to
 the transition form factors in Eq.~(\ref{formsr}) are given here.
 The contributions of condensate of dimension-six operator
 $\langle \bar{\Psi}\Psi \rangle ^2$ are numerically negligible,
 whereas their expressions are more tedious, therefore we do not
 present all of them here.

\begin{footnotesize}
\vspace{1cm}

\noindent 1) Results for $f_0$:

$$ \hat{B}f_0= \hat{B}f_0^{pert}+ \hat{B}f_0^{(3)}+
\hat{B}f_0^{(5)}+ \hat{B}f_0^{(6)}\; ,$$
with
$$
\begin{array}{ll}
 \hat{B}f_0^{pert}~=&\int^{s_2^0}_{4m_s^2}
 ds_2\int^{s_1^0}_{s_1^L}ds_1
\displaystyle
\frac{3e^{-s_1/M_1^2-s_2/M_2^2}}{4M_1^2M_2^2\pi^2\lambda^{3/2}}
 [2s_2m_c^3-2s_2m_sm_c^2-s_2(2m_s^2\\[4mm]
  &+s_1-s_2+q^2)m_c
+m_s(-s_2^2+2m_s^2s_2+s_1s_2+q^2s_2+\lambda)],
\end{array}
\eqno(A1)$$
where $\lambda=(s_1+s_2-q^2)^2-4s_1 s_2$. The lower integration limit
$s_1^L$  is determined by the condition that all internal quarks
are on their mass shell \cite{Landau},
$$ s_1^L=\frac{m_c^2}{m_c^2-q^2}s_2+m_c^2\; .$$

$$
\begin{array}{ll}
 \hat{B}f_0^{(3)}~=&-\displaystyle
\frac{e^{-m_c^2/M_1^2-m_s^2/M_2^2}}{6M_1^8M_2^8}[(6M_1^2M_2^6-3(M_1^2+q^2)m_s^2M_2^4
  +(4(M_1^2+M_2^2)\\[4mm]
&+q^2)m_s^4M_2^2-(M_1^2+M_2^2)m_s^6)M_1^4+M_2^2(M_1^2+M_2^2)m_c^2m_s^2(3M_2^2\\[4mm]
&-m_s^2)M_1^2+M_2^2m_cm_s(-3M_1^2M_2^4
+(M_1^2+M_2^2+q^2)m_s^2M_2^2\\[4mm]
&-(M_1^2+M_2^2)m_s^4)M_1^2 -M_2^4(M_1^2+M_2^2)m_c^3m_s^3]
\times\langle \bar{s}s\rangle\; ,
 \end{array}
\eqno(A2)
$$

$$
\begin{array}{ll}
 \hat{B}f_0^{(5)}~=&-\displaystyle
\frac{e^{-m_c^2/M_1^2-m_s^2/M_2^2}}{12M_1^8M_2^8}[(3M_2^6-M_1^2M_2^4+(M_1^2+M_2^2)m_s^4\\[4mm]
&-(3M_2^4+5M_1^2M_2^2)m_s^2)M_1^4-M_2^2(M_1^2+M_2^2)m_c^2(3M_2^2-m_s^2)M_1^2\\[4mm]
&+M_2^2m_cm_s(2M_2^4-2M_1^2M_2^2+(M_1^2+M_2^2)m_s^2)M_1^2+M_2^4(M_1^2+M_2^2)m_c^3m_s\\[4mm]
&+q^2(M_1^4M_2^2(3M_2^2-m_s^2)-M_1^2M_2^4m_cm_s)]\times g\langle
\bar{s}\sigma TG s \rangle\; ,
\end{array}
\eqno(A3)
$$

$$
\begin{array}{ll}
 \hat{B}f_0^{(6)}~=&\displaystyle
\frac{e^{-m_c^2/M_1^2-m_s^2/M_2^2}}{81M_1^8M_2^8(m_c^2-q^2)m_s^3}
[18(-1+e^{m_s^2/M_2^2})M_1^6(2m_c-m_s)m_sM_2^6\\[4mm]
&+(M_1^2+M_2^2)m_c^5m_s^3M_2^4+M_1^2(M_1^2+M_2^2)m_c^4m_s^4M_2^2\\[4mm]
&+M_1^2q^4m_s^3(m_sM_1^2+M_2^2m_c)M_2^2+M_1^2m_c^3m_s^3(-13M_2^4+2M_1^2M_2^2\\[4mm]
&+(M_1^2+M_2^2)m_s^2)M_2^2+M_1^4m_c^2(-54(-1+e^{m_s^2/M_2^2})M_1^2M_2^6\\[4mm]
&+54M_1^2m_s^2M_2^4-(M_1^2+10M_2^2)m_s^4M_2^2+(M_1^2+M_2^2)m_s^6)\\[4mm]
&+q^2(54(-1+e^{m_s^2/M_2^2})M_2^6M_1^6-54M_2^4m_s^2M_1^6-(M_1^2+M_2^2)m_s^6M_1^4\\[4mm]
&-M_2^2(M_1^2+M_2^2)m_cm_s^5M_1^2+M_2^2(M_1^4+10M_2^2M_1^2-(2M_1^2\\[4mm]
&+M_2^2)m_c^2)m_s^4M_1^2-M_2^2m_c(2M_1^4-13M_2^2M_1^2+(2M_1^2\\[4mm]
&+M_2^2)m_c^2)m_s^3)] \times g^2\langle \bar{s}s\rangle^2\; .
\end{array}
\eqno(A4)
$$

\vspace{1cm}

\noindent 2) Results for $f_1+f_3$ :

$$ \hat{B}(f_1+f_3)= \hat{B}f_+^{pert}+ \hat{B}f_+^{(3)}+
\hat{B}f_+^{(5)}+ \hat{B}f_+^{(6)}\; ,$$
 with
$$
\begin{array}{ll}
 \hat{B}f_+^{pert}~=&\int^{s_2^0}_{4m_s^2} ds_2\int^{s_1^0}_{s_1^L}
\displaystyle
\frac{3e^{-s_1/M_1^2-s_2/M_2^2}}{4M_1^2M_2^2\pi^2\lambda^{5/2}}
\{-6s_2(-s_1+s_2+q^2)m_c^5+6s_2(-s_1\\[4mm]
&+s_2+q^2)m_sm_c^4+2s_2(-4s_1^2+8s_2s_1-4s_2^2+2q^4-6(s_1-s_2)m_s^2\\[4mm]
&+\lambda+2q^2(3m_s^2+s_1+s_2))m_c^3+2s_2m_s[4s_1^2-8s_2s_1+4s_2^2-2q^4\\[4mm]
&+6(s_1-s_2)m_s^2-3\lambda-2q^2(3m_s^2+s_1+s_2)]m_c^2+[6(s_1\\[4mm]
&-s_2)s_2m_s^4+2(4s_2^3-8s_1s_2^2+4s_1^2s_2-2\lambda s_2+s_1\lambda)m_s^2+(s_1\\[4mm]
&-s_2)s_2(2s_1^2-4s_2s_1+2s_2^2-\lambda)-2s_2q^4(2m_s^2+2s_1\\[4mm]
&+s_2)+q^2(-6s_2m_s^4-2(2s_2^2+2s_1s_2+\lambda)m_s^2+s_2(2s_1^2-6s_2s_1\\[4mm]
&+4s_2^2-\lambda))]m_c+m_s[2s_2^4-6s_1s_2^3+6s_1^2s_2^2-3\lambda
s_2^2+6(s_2\\[4mm]
&-s_1)m_s^4s_2-s_1^3s_2+3s_1\lambda
s_2+2q^4(2m_s^2+2s_1+s_2)s_2+\lambda^2\\[4mm]
&-2(4s_2^3-8s_1s_2^2+4s_1^2s_2-4\lambda s_2+s_1\lambda
)m_s^2+q^2(6s_2m_s^4+2(2s_2^2\\[4mm]
&+2s_1s_2+\lambda )m_s^2+s_2(-2s_1^2+6s_2s_1-4s_2^2+3\lambda
))]\}\; ,
\end{array}
\eqno(A5)
$$

$$
\begin{array}{ll}
 \hat{B}f_+^{(3)}~=&-\displaystyle
\frac{e^{-m_c^2/M_1^2-m_s^2/M_2^2}}{6M_1^8M_2^8}\{[6M_1^2M_2^6
-3(M_1^2-2M_2^2+q^2)m_s^2M_2^4+(4(M_1^2\\[4mm]
&+M_2^2)+q^2)m_s^4M_2^2-(M_1^2+M_2^2)m_s^6]M_1^4+M_2^2(M_1^2+M_2^2)m_c^2m_s^2(3M_2^2\\[4mm]
&-m_s^2)M_1^2-M_2^2m_cm_s(3M_1^2M_2^4+(M_1^2+M_2^2-q^2)m_s^2M_2^2\\[4mm]
& +(M_1^2+M_2^2)m_s^4)M_1^2-M_2^4(M_1^2+M_2^2)m_c^3m_s^3 \}
\times\langle \bar{s}s\rangle\; ,
 \end{array}
\eqno(A6)
$$

$$
\begin{array}{ll}
 \hat{B}f_+^{(5)}~=&\displaystyle
\frac{e^{-m_c^2/M_1^2-m_s^2/M_2^2}}{12M_1^8M_2^8}\{
[(M_1^2+3M_2^2)M_2^4-(M_1^2+M_2^2)m_s^4+(7M_2^4\\[4mm]
&+5M_1^2M_2^2)m_s^2]M_1^4
+M_2^2(M_1^2+M_2^2)m_c^2(3M_2^2-m_s^2)M_1^2-M_2^2m_cm_s\\[4mm]
&[2(M_1^2+2M_2^2)M_2^2+(M_1^2+M_2^2)m_s^2
]M_1^2-M_2^4(M_1^2+M_2^2)m_c^3m_s+q^2[M_2^2(m_s^2\\[4mm]
&-3M_2^2)M_1^4
+M_2^4m_cm_sM_1^2]\}\times g\langle
\bar{s}\sigma TG s \rangle\; ,
\end{array}
\eqno(A8)
$$
$$
\begin{array}{ll}
 \hat{B}f_+^{(6)}~=&\displaystyle
\frac{e^{-m_c^2/M_1^2-m_s^2/M_2^2}}{81M_1^8M_2^8(m_c^2-q^2)m_s^3}
\{-18(-1+e^{m_s^2/M_2^2})M_1^6m_s^2M_2^6+(7M_2^2\\[4mm]
&-2M_1^2)m_c^5m_s^3M_2^4
+M_1^2(M_1^2+M_2^2)m_c^4m_s^4M_2^2+M_1^2q^4m_s^3(M_1^2m_s\\[4mm]
&-2M_2^2m_c)M_2^2+
M_1^2m_c^3m_s^3[-26M_2^4+4M_1^2M_2^2+(M_1-2M_2^2)m_s^2]\\[4mm]
&M_2^2+
M_1^4m_c^2[-54(-1+e^{m_s^2/M_2^2})M_1^2M_2^6+54M_1^2m_s^2M_2^4
-4(M_1^2\\[4mm]
&-2M_2^2)m_s^4M_2^2+(M_1^2+M_2^2)m_s^6]
+q^2[54(-1+e^{m_s^2/M_2^2})M_1^6M_2^6\\[4mm]
&-54M_2^4m_s^2M_1^6
-(M_1^2+M_2^2)m_s^6M_1^4-M_2^2(M_1^2-2M_2^2)m_cm_s^5M_1^2\\[4mm]
&+M_2^2(4(M_1^4-2M_1^2M_2^2)-(2M_1^2+M_2^2)m_c^2)m_s^4M_1^2
+M_2^4m_c\\[4mm]
&(-4M_1^4+26M_2^2M_1^2+(4M_1^2-7M_2^2)m_c^2)m_s^3]\}
 \times g^2\langle \bar{s}s\rangle^2\; .
\end{array}
\eqno(A9)
$$
\vspace{1cm}

\noindent 3) Results for $f_1-f_3$ :

$$ \hat{B}(f_1-f_3)= \hat{B}f_-^{pert}+ \hat{B}f_-^{(3)}+
\hat{B}f_-^{(5)}+ \hat{B}f_-^{(6)}\; ,$$
 with
$$
\begin{array}{ll}
 \hat{B}f_-^{pert}~=&\int^{s_2^0}_{4m_s^2} ds_2\int^{s_1^0}_{s_1^L}
\displaystyle
\frac{-3e^{-s_1/M_1^2-s_2/M_2^2}}{4M_1^2M_2^2\pi^2\lambda^{5/2}}
\{6s_2(s_1+3s_2-q^2)m_c^5-6s_2(s_1\\[4mm]
&+3s_2-q^2)m_sm_c^4
+2s_2[-4s_1^2-4s_2s_1+8s_2^2+2q^4-6(s_1+3s_2)\\[4mm]
&m_s^2+\lambda
+2q^2(3m_s^2+s_1-5s_2)]m_c^3+2s_2m_s[4s_1^2+4s_2s_1-8s_2^2\\[4mm]
&-2q^4
+6(s_1+3s_2)m_s^2+\lambda -2q^2(3m_s^2+s_1-5s_2)]m_c^2
+[6s_2(s_1\\[4mm]
&+3s_2)m_s^4+2(s_1+2s_2)(-4s_2^2+4s_1s_2+\lambda )m_s^2
+(s_1-s_2)s_2(2s_1^2\\[4mm]
&-2s_2^2-\lambda )+2s_2 q^4(-2m_s^2-2s_1+s_2)
+q^2(-6s_2m_s^4-2(-10s_2^2\\[4mm]
&+2s_1s_2+\lambda
)m_s^2-s_2(-2s_1^2-10s_2s_1+4s_2^2+\lambda))]m_c
-m_s[2s_2^4\\[4mm]
&-2s_1s_2^3-2s_1^2s_2^2-\lambda
s_2^2+6(s_1+3s_2)m_s^4s_2+2s_1^3s_2+s_1\lambda s_2\\[4mm]
&+2q^4(-2m_s^2-2s_1+s_2)s_2-\lambda^2+2(-8s_2^3+4s_1s_2^2+4s_1^2s_2\\[4mm]
&+4\lambda s_2+s_1\lambda )m_s^2
+q^2(-6s_2m_s^4-2(-10s_2^2+2s_1s_2+\lambda )m_s^2\\[4mm]
&+s_2(2s_1^2+10s_2s_1-4s_2^2+\lambda ))]\}\; ,
\end{array}
\eqno(A10)
$$
$$
\begin{array}{ll}
 \hat{B}f_-^{(3)}~=&\displaystyle
\frac{e^{-m_c^2/M_1^2-m_s^2/M_2^2}}{6M_1^8M_2^8}\{[6M_1^2M_2^6
-3(M_1^2+2M_2^2+q^2)m_s^2M_2^4+(4(M_1^2\\[4mm]
&+M_2^2)+q^2)m_s^4M_2^2-(M_1^2+M_2^2)m_s^6]M_1^4+M_2^2(M_1^2+M_2^2)m_c^2\\[4mm]
&m_s^2(3M_2^2-m_s^2)M_1^2-M_2^2m_cm_s(3M_1^2M_2^4+(M_1^2-3M_2^2-q^2)m_s^2\\[4mm]
& M_2^2+(M_1^2+M_2^2)m_s^4)M_1^2-M_2^4(M_1^2+M_2^2)m_c^3m_s^3 \}
\times\langle \bar{s}s\rangle\; ,
 \end{array}
\eqno(A11)
$$

$$
\begin{array}{ll}
 \hat{B}f_-^{(5)}~=&-\displaystyle
\frac{e^{-m_c^2/M_1^2-m_s^2/M_2^2}}{12M_1^8M_2^8}\{
[(M_1^2-9M_2^2)M_2^4-(M_1^2+M_2^2)m_s^4+(5M_1^2M_2^2\\[4mm]
&-M_2^4)m_s^2]M_1^4
+M_2^2(M_1^2+M_2^2)m_c^2(3M_2^2-m_s^2)M_1^2\\[4mm]
&-M_2^4(M_1^2+M_2^2)m_c^3m_s-m_c[2M_1^4m_sM_2^4+M_1^2(M_1^2+M_2^2)m_s^3M_2^2
]\\[4mm]
&+q^2[M_2^2(m_s^2-3M_2^2)M_1^4
+M_2^4m_cm_sM_1^2]\}\times g\langle
\bar{s}\sigma TG s \rangle\; ,
\end{array}
\eqno(A13)
$$
$$
\begin{array}{ll}
 \hat{B}f_-^{(6)}~=&-\displaystyle
\frac{e^{-m_c^2/M_1^2-m_s^2/M_2^2}}{81M_1^8M_2^8(m_c^2-q^2)m_s^3}
\{54(-1+e^{m_s^2/M_2^2})M_1^6m_s^2M_2^6+(7M_2^2\\[4mm]
&-2M_1^2)m_c^5m_s^3M_2^4
+M_1^2(M_1^2+M_2^2)m_c^4m_s^4M_2^2+M_1^2q^4m_s^3(M_1^2m_s\\[4mm]
&-2M_2^2m_c)M_2^2+
M_1^2m_c^3m_s^3[-18M_2^4+4M_1^2M_2^2+(M_1-2M_2^2)m_s^2]\\[4mm]
&M_2^2+
M_1^4m_c^2[-54(-1+e^{m_s^2/M_2^2})M_1^2M_2^6+54M_1^2m_s^2M_2^4
-4(M_1^2\\[4mm]
&+4M_2^2)m_s^4M_2^2+(M_1^2+M_2^2)m_s^6]
+q^2[54(-1+e^{m_s^2/M_2^2})M_1^6M_2^6\\[4mm]
&-54M_2^4m_s^2M_1^6
-(M_1^2+M_2^2)m_s^6M_1^4-M_2^2(M_1^2-2M_2^2)m_cm_s^5M_1^2\\[4mm]
&+M_2^2(4M_1^2(M_1^2+4M_2^2)-(2M_1^2+M_2^2)m_c^2)m_s^4M_1^2
+M_2^4m_c\\[4mm]
&(-4M_1^4+18M_2^2M_1^2+(4M_1^2-7M_2^2)m_c^2)m_s^3]\}
 \times g^2\langle \bar{s}s\rangle^2\; .
\end{array}
\eqno(A14)
$$
\vspace{1cm}

\noindent 4) Results for $f_5$ :

$$ \hat{B}(f_5)= \hat{B}f_5^{pert}+ \hat{B}f_5^{(3)}+
\hat{B}f_5^{(5)}+ \hat{B}f_5^{(6)}\; ,$$
 with
$$
\begin{array}{ll}
 \hat{B}f_5^{pert}~=&\int^{s_2^0}_{4m_s^2} ds_2\int^{s_1^0}_{s_1^L}
\displaystyle
\frac{-3e^{-s_1/M_1^2-s_2/M_2^2}}{8M_1^2M_2^2\pi^2\lambda^{3/2}}
\{2s_2m_c^5-2s_2m_sm_c^4-2s_2(2m_s^2+s_1\\[4mm]
&-s_2+q^2)m_c^3+2s_2m_s(2M-s^2
+s_1-s_2+q^2)m_c^2+[2s_2m_s^4\\[4mm]
&+2(-s_2^2+s_1s_2+\lambda )m_s^2
+s_2\lambda +2s_2q^2
(m_s^2+s_1)]m_c-m_s[2s_2m_s^4\\[4mm]
&+2(-s_2^2+s_1s_2+\lambda )m_s^2
-s_1\lambda +q^2(2s_2m_s^2+2s_1s_2+\lambda )]\} \; ,
\end{array}
\eqno(A15)
$$
$$
\begin{array}{ll}
 \hat{B}f_5^{(3)}~=&-\displaystyle
\frac{e^{-m_c^2/M_1^2-m_s^2/M_2^2}}{12M_1^8M_2^8}\{M_2^4(M_1^2+M_2^2)m_s^3m_c^5
-M_2^2(M_1^2+M_2^2)m_s^2\\[4mm]
&[3M_1^2M_2 -(M_1^2+2M_2^2)m_s^2]m_c^4
+[3M_1^4m_sM_2^6-9M_1^2(M_1^2+M_2^2)\\[4mm]
&m_s^3M_2^4+(M_2^6+4M_1^2M_2^4+3M_1^4M_2^2)m_s^5]
m_c^3+M_1^2[-6M_1^4M_2^6+(M_1^4\\[4mm]
&+4M_2^2M_1^2+3M_2^4)m_s^6-(6M_2^6+11M_1^2M_2^4
+5M_1^4M_2^2)m_s^4\\[4mm]
&+3(3M_1^2M_2^6+2M_1^4M_2^4)m_s^2]m_c^2+M_1^2m_s[-15M_1^4M_2^6
+(2M_1^4\\[4mm]
&+3M_2^2M_1^2+M_2^4)m_s^6-(2M_2^6+13M_1^2M_2^4+11M_1^4M_2^2)m_s^4\\[4mm]
&+(8M_1^2M_2^6
+11M_1^4M_2^4)m_s^2]m_c+M_1^4m_s^2[3M_1^2M_2^6-4(M_1^2+M_2^2)\\[4mm]
&m_s^4M_2^2+(M_1^2+M_2^2)m_s^6
+(2M_2^6+5M_1^2M_2^4)m_s^2]+M_1^2M_2^2q^4m_s^2\\[4mm]
&[(m_s^2-3M_2^2)M_1^2+M_2^2m_cm_s]+
q^2[(6M_1^2M_2^6-(M_1^2+2M_2^2)m_s^6\\[4mm]
&+(7M_2^4+5M_1^2M_2^2)m_s^4-3(M_2^6+2M_1^2M_2^4)m_s^2)M_1^4
+M_2^2m_c^2m_s^2\\[4mm]
&(3M_2^2(2M_1^2+M_2^2)-(2M_1^2+3M_2^2)m_s^2)M_1^2+M_2^2m_cm_s
(-3M_1^2M_2^4\\[4mm]
&-(3M_1^2+2M_2^2)m_s^4+(2M_2^4+9M_1^2M_2^2)m_s^2)M_1^2-M_2^4(2M_1^2+M_2^2)\\[4mm]
&m_c^3m_s^3]\}
\times\langle \bar{s}s\rangle\; ,
 \end{array}
\eqno(A16)
$$

$$
\begin{array}{ll}
 \hat{B}f_5^{(5)}~=&\displaystyle
\frac{e^{-m_c^2/M_1^2-m_s^2/M_2^2}}{24M_1^8M_2^8}\{
M_2^4(M_1^2+M_2^2)m_sm_c^5-M_2^2(M_1^2+M_2^2)[3M_1^2M_2^2\\[4mm]
&-(M_1^2+2M_2^2)m_s^2]m_c^4
+M_2^2m_s[(3M_1^4+4M_2^2M_1^2+M_2^4)m_s^2\\[4mm]
&-2M_1^2M_2^2(5M_1^2+3M_2^2)]m_c^3
+[2M_1^4(M_1^2+3M_2^2)M_2^4+(M_1^6\\[4mm]
&+4M_2^2M_1^4+3M_2^4M_1^2)m_s^4-
2(3M_2^2M_1^6+5M_2^4M_1^4)m_s^2]m_c^2+M_1^2m_s\\[4mm]
&[4M_1^2(M_1^2+3M_2^2)M_2^4+2(M_1^4
+3M_2^2M_1^2+M_2^4)m_s^4+(M_2^6\\[4mm]
&-8M_1^2M_2^4-13M_1^4M_2^2)m_s^2]m_c
+M_1^4[4M_1^2M_2^6+(M_1^2+M_2^2)\\[4mm]
&m_s^2(2M_2^4+m_s^4)-(M_2^4+5M_1^2M_2^2)m_s^4]+q^4[M_2^2(m_s^2\\[4mm]
&-3M_2^2)M_1^4+M_2^4m_cm_sM_1^2]-q^2[(2(M_1^2+3M_2^2)M_2^4
+(M_1^2+2M_2^2)m_s^4\\[4mm]
&-2(2M_2^4+3M_1^2M_2^2)m_s^2)M_1^4+M_2^2m_cm_s(M_2^4
-10M_1^2M_2^2\\[4mm]
&+(3M_1^2+2M_2^2)m_s^2)M_1^2+M_2^2m_c^2((2M_1^2+3M_2^2)m_s^2\\[4mm]
&-3(M_2^4+2M_1^2M_2^2))M_1^2+M_2^4(2M_1^2+M_2^2)m_c^3m_s]
\}\times g\langle
\bar{s}\sigma TG s \rangle\; .
\end{array}
\eqno(A18)
$$
\end{footnotesize}

\end{document}